\documentclass[journal,11pt,draftclsnofoot,onecolumn]{IEEEtran}

\IEEEoverridecommandlockouts                              

\usepackage{graphicx}
\usepackage[square,comma,numbers,sort&compress]{natbib}
\usepackage{amsthm,amssymb,amscd,amssymb,amsmath,mathrsfs}
\usepackage{epstopdf}
\usepackage{balance}
\usepackage[noend]{algorithm} 
\usepackage{algorithmicx} 
\usepackage[noend]{algpseudocode}
\usepackage{multirow} 
\usepackage{amsmath}
\usepackage{xcolor}
\usepackage{lineno}
\usepackage{wrapfig}
\usepackage{enumerate}
\usepackage{changepage}

\theoremstyle{plain}
\newtheorem{theorem}{Theorem}

\newtheorem{lemma}{Lemma}

\theoremstyle{definition}
\newtheorem{definition}{Definition}

\theoremstyle{remark}
\newtheorem{remark}{Remark}

\begin{document}
%
\title{Supervisor Localization of Timed Discrete-Event Systems
under Partial Observation and Communication Delay}

\author{Renyuan Zhang$^{1}$ and Kai Cai$^{2}$
\thanks{*This work was supported in part by the National Nature Science Foundation of China, Grant no. 61573289, 11772264;
JSPS KAKENHI Grant no. JP16K18122.
}
\thanks{$^{1}$R. Zhang is with School of Automation, Northwestern Polytechnical University, China
        {\tt\small ryzhang@nwpu.edu.cn}}%
\thanks{$^{2}$K. Cai is with Department of Electrical and Information Engineering, Osaka City University, Japan
        {\tt\small kai.cai@eng.osaka-cu.ac.jp}}%
}

\maketitle


\thispagestyle{empty} \pagestyle{plain}

\begin{abstract}
We study \emph{supervisor localization} for timed discrete-event systems
under partial observation and communication delay in the Brandin-Wonham
framework. First, we employ timed relative observability to synthesize
a partial-observation monolithic supervisor; the control actions of this
supervisor include not only disabling action of prohibitible events
(as that of controllable events in the untimed case) but also
``clock-preempting'' action of forcible events. Accordingly we decompose
the supervisor into a set of partial-observation local controllers one for
each prohibitible event, as well as a set of partial-observation local
preemptors one for each forcible event. We prove that these local controllers
and preemptors collectively achieve the same controlled behavior as the
partial-observation monolithic supervisor does. Moreover, we propose
channel models for inter-agent event communication with bounded and unbounded
delays; the channel models are treated as plant components. In this formulation, there
exist multiple distinct observable event sets; thus we employ timed
relative coobservability to synthesize partial-observation decentralized
supervisors, and then localize these supervisors into local controllers and
preemptors. The above results are illustrated by a timed workcell example.
\end{abstract}


\section{Introduction}


In \cite{CaiWon10a,CaiWon10b,CaiWon16}
we developed a top-down approach, called {\it supervisor localization}, to the
distributed control synthesis of multi-component 
discrete-event systems
(DES). The essence of localization is the decomposition of the
monolithic (optimal and nonblocking) supervisor into local controllers
for the individual components. 
In \cite{ZhangEt13} we extended
supervisor localization to timed DES (TDES) in the Brandin-Wonham
framework \cite{BraWon94}; in addition to local controllers
(corresponding to disabling actions), a set of local preemptors is
obtained corresponding to clock-preempting actions. More recently
in \cite{ZhangCW17}, we extended the untimed supervisor
localization to the case of partial observation. In particular,
we combined localization with relative observability
\cite{CaiZW15} to first synthesize a partial-observation
monolithic supervisor, and then decompose the supervisor into
local controllers whose state changes are caused only by observable
events.

In this paper and its conference precursor \cite{ZhangCai16b},
we generalize supervisor localization to study distributed control of multi-component 
 TDES under partial observation and communication
delay. Our study is divided into two parts.
In the first part, we focus on partial-observation supervisor
localization for TDES in the Brandin-Wonham framework, thereby
extending both \cite{ZhangEt13} and \cite{ZhangCW17}.
We propose to first synthesize a partial-observation monolithic
supervisor using the concept of timed relative observability
\cite{CaiZW16}. Timed relative observability is proved to be
generally stronger than timed observability \cite{LinWon95},
weaker than normality \cite{LinWon95}, and closed under set
union. Therefore the supremal timed relatively observable (and controllable)
sublanguage of a given language exists and may be effectively computed
\cite{CaiZW16}. Since this supremal sublanguage is timed observable
and controllable, it may be implemented by a partial-observation
(feasible and nonblocking) supervisor \cite{LinWon95}. We then
suitably extend the localization procedure in \cite{ZhangEt13}
to decompose the supervisor into partial-observation local controllers
and local preemptors for individual components, 
and prove that the derived
local controlled behavior is equivalent to the monolithic one
and is therefore globally observable and controllable.

In the second part, we consider not only partial observation, but also
that inter-agent\footnote{We view that an {\it agent} is a plant component equipped with
a set of partial-observation local controllers/preemptors.} event communication is subject to delay. First, we
introduce two types of channel models for inter-agent event communication.
The introduced models are treated as plant components. In this formulation,
the observable event sets of different agents are generally distinct. This is because the
occurrence of a communication event and sending that event are observable
only to the sender, but not observable to the receiver; on the other hand,
receiving of a communication event is observable only to the receiver,
but not observable to the sender. To deal with multiple observable event
sets, we propose to employ the concept of timed relative coobservability
\cite{CaiZW16}, which is closed under set union, to first synthesize
a set of partial-observation decentralized supervisors, and then decompose
these decentralized supervisors into the respective local controllers/preemptors.
Finally, we prove that the derived local controlled behavior is identical
to that achieved by the partial-observation decentralized supervisors.

The main contributions of this work are as follows.

1) The proposed timed
supervisor localization under partial observation extends the untimed
counterpart in \cite{ZhangCW17} and the full-observation counterpart in \cite{ZhangEt13}.
Compared with \cite{ZhangCW17}, not only is the monolithic supervisor's
disabling action localized (as in the untimed case), but also its preemptive
action is localized with respect to individual forcible events. While compared
with \cite{ZhangEt13}, the new concepts of {\it partial-observation control cover} and
{\it partial-observation preemption cover} are defined on the {\it powerset} of the monolithic supervisor's state set. In this
way, in the transition structure of the resulting local controllers/preemptors,
only observable events can lead to state changes. It is important to stress that
the proposed timed supervisor localization under partial observation cannot be
obtained directly by combining \cite{ZhangEt13} and \cite{ZhangCW17}, because the treatment of event $tick$ is new and
cannot be found in \cite{ZhangEt13} or \cite{ZhangCW17}:
\begin{itemize}
\item The partial-observation local preemptor (${\bf LOC}_\alpha^P$ in
Section~\ref{sec:ProbFormu}) accounts for both the effect of partial observation and localization
of $tick$-preempting actions to individual forcible events.

\item The $tick$-enabling function and $tick$-preemption
function ($E_{tick}$ and $F_\alpha$ in Section~\ref{subsec:PreemptAct}) are defined using the state
set and transition function of the monolithic supervisor,
as well as the set of uncertainty (state) sets caused by partial
observation and the associated transition function.

\item Two new functions ($\psi_\alpha$ and $\psi_{tick}$ in Section~\ref{subsec:PreemptAct}) need
to be defined from the partial-observation preemption cover, so that
certain unobservable events (possibly including $tick$) can be appropriately
added as selfloops.

\item It is established (Theorem 1 in Section~\ref{subsec:mainresult}) that the resulting
partial-observation local $tick$-preemptors and local controllers
    collectively achieve the monolithic $tick$-preemption and disabling controlled
    behavior.

\end{itemize}

2) In addition to partial
observation, timed supervisor localization is extended to address communication
delay.

\begin{itemize}
\item A TDES channel model (Section~\ref{subsec:chnmodel}) which can represent bounded and unbounded
communication delays is adopted. Unlike \cite{ZhangEt16a,ZhangEt16b}, the channel model is treated as plant component,
and thus the communication delays are integrated into the plant behavior.

\item Timed relative coobservability \cite{CaiZW16} is adopted to
effectively compute decentralized supervisors tolerant of communication delays.
Relative coobservability is stronger than coobservability, but permits existence
of the supremal element; an algorithm in \cite{CaiZW16} effectively computes the supremal relatively coobservable
sublanguage of a given (non-closed) language. The combination of timed
relative coobservability and partial-observation supervisor localization
is new, and leads to a computationally effective solution to delay-tolerant
distributed control.

\item  It is established (Theorem 2 in Section~\ref{subsec:delaymainresult}) that the resulting
partial-observation local $tick$-preemptors and local controllers collectively
tolerate specified bounded and unbounded communication delays.
\end{itemize}

Overall, the proposed supervisor localization for TDES provides a top-down, computationally
effective approach to the distributed control of timed DES under partial observation and
communication delay, which was not available in the literature. By the allocation policy
described in Section \ref{sec:ProbFormu}, the partial-observation local preemptors/controllers
derived by the proposed localization procedures are allocated to each plant component,
thereby building a purely distributed control architecture.

We note in the literature that the
algorithms in \cite{TakUsh03,YinLaf16a} compute a nonblocking
(maximally) observable sublanguage that is generally incomparable
with the supremal relatively observable sublanguage. The reason that we adopt
relative observability/coobservability is first of all that their generator-based
computations of the supremal sublanguages are better suited for applying our
localization algorithm whose computations are also generator-based; together
they constitute a computationally effective synthesis approach.
 Another important reason is that when introducing inter-agent communication,
individual agents may have distinct observable event sets, and in this situation,
relative coobservability is essential to compute a set of partial-observation decentralized supervisors
(for the given specification language is generally not coobservable).
It is interesting to explore the combinations of partial-observation localization procedure
with the algorithms in \cite{TakUsh03,YinLaf16a}; we shall leave this for our future work.

We note also that distributed/decentralized supervisory control with communication
delay has been extensively studied. First, to capture communication delays in multi-component
plant, there are mainly two approaches reported in the literature. The first is to model the
communication by separate models, e.g. information structure \cite{BarLaf00},
FIFO queue \cite{Tripak04,Hirais09,KalyoEt11}, shared medium communication model \cite{Schmid07};
then the plant behavior with communication delay will be obtained through appropriate
composition operators on the plant components and the communication models.
The other approach is to define observation maps \cite{ParCho07,XuKum08,RicCai11,Lin14} on
the plant behavior; then the plant behavior with communication delay is exactly the codomain of
the observation maps. In this paper, we used a TDES channel model in which
the communication delays are measured by number of $tick$s, and the delays at each transmission period
are modeled separately. Compared with the models in the literature, our channel model is represented
by TDES and treated as plant component; thus the plant behavior with delay
can be obtained by synchronous product defined on (generalized) TDES, rather than by any newly defined composition operators.


Second, to synthesize distributed/decentralized supervisors that are able to tolerate specified communication delays,
there are mainly two approaches reported in the literature.
The first is a verification approach, e.g. \cite{ZhangEt16a,ZhangEt16b, SadiRH15},
which first synthesizes delay-free distributed controllers,
and then verifies whether the distributed controllers tolerate given communication
delays. This approach is limited to verifying the robustness of derived controllers \cite{ZhangEt16a,ZhangEt16b}
or that of existing communication protocols \cite{SadiRH15},
but does not supply a procedure to construct controllers that are able to tolerate
given communication delays.
The second approach is that of synthesis, e.g.
\cite{BarLaf00, Tripak04, ParCho07, Hirais09, RicCai11, Lin14}, which first incorporates
communication delays into the plant and specification models, and then applies
decentralized control methods to synthesize distributed/decentralized controllers that tolerate
the communication delay. In these works, {\it observability} \cite{BarLaf00,Hirais09},
{\it joint observability} \cite{Tripak04}, {\it coobservability} \cite{RicCai11},
{\it delay-coobservability} \cite{ParCho07}, or {\it network observability}
\cite{Lin14} are necessary for the existence of distributed controllers.
However, these observability properties are not closed under set union, and thus
there generally does not exist the respective supremal sublanguage of a given language.
By contrast, we employ the recently proposed timed relative coobservability, which is closed
under set union and the supremal relatively coobservable sublanguage is effectively
computable \cite{CaiZW16}. Other issues including state
avoidance control problem with communication delay \cite{KalyoEt11}, delay effects in implementation
of decentralized/distributed supervisors \cite{Schmid07,XuKum08},  are also reported in the literature;
we refer to \cite{ZhangEt16a,ZhangEt16b} for a detailed review.

The paper is organized as follows. Section~\ref{sec:prelim} reviews the preliminaries on the
Brandin-Wonham  TDES framework. Section~\ref{sec:ProbFormu} formulates the partial-observation
supervisor localization problem of TDES, and Section~\ref{sec:LocProc} develops the solution
localization procedure. Section~\ref{sec:delay} investigates partial-observation supervisor
localization with communication delay by using the concept of timed relative coobservability.
Finally Section~\ref{sec:concl} states our conclusions.


\section{Preliminaries} \label{sec:prelim}


This section reviews supervisory control of TDES
in the Brandin-Wonham framework \cite{BraWon94},\cite[Chapter~9]{Wonham16a}.
First consider the untimed DES model ${\bf G}_{act} = (A, \Sigma_{act}, \delta_{act}, a_0, A_m)$;
here $A$ is the finite set of {\it activities}, $\Sigma_{act}$
the finite set of {\it events}, $\delta_{act}:A \times \Sigma_{act}
\to A$ the (partial) {\it transition function}, $a_0 \in
A$ the {\it initial activity}, and $A_m \subseteq A$ the set
of {\it marker activities}. Let $\mathbb{N}$ denote the set of
natural numbers $\{0,1,2,...\}$, and introduce \emph{time} into ${\bf
G}_{act}$ by assigning to each event $\sigma \in \Sigma_{act}$ a
{\it lower bound} $l_{{\bf G},\sigma} \in \mathbb{N}$ and an {\it upper
bound} $u_{{\bf G},\sigma} \in \mathbb{N} \cup\{\infty\}$, such that
$l_{{\bf G},\sigma} \leq u_{{\bf G},\sigma}$.
Also introduce a distinguished event, written $tick$, to
represent ``tick of the global clock''. Then a TDES model
\begin{equation} \label{eq:G}
{\bf G} := (Q, \Sigma, \delta, q_0, Q_m),
\end{equation}
is constructed from ${\bf G}_{act}$ (refer to \cite{BraWon94},
\cite[Chapter~9]{Wonham16a} for detailed construction)
such that $Q$ is the finite set of
\emph{states}, $\Sigma := \Sigma_{act} \dot\cup \{tick\}$
the finite set of events, $\delta:Q\times\Sigma \rightarrow Q$
the (partial) {\it state transition function}, $q_0$ the {\it
initial state}, and $Q_m$ the set of {\it marker states}.

Let $\Sigma^*$ be the set of all finite strings of elements in
$\Sigma = \Sigma_{act} \dot\cup \{tick\}$, including the empty
string $\epsilon$.  The transition function $\delta$ is extended to
$\delta:Q\times \Sigma^* \rightarrow Q$ in the usual way. The {\it
closed behavior} of $\bf G$ is the language 
$L({\bf G}) := \{s \in \Sigma^*|\delta(q_0,s)!\}$
and the {\it marked behavior} is 
$L_m({\bf G}) := \{s \in L({\bf G})| \delta(q_0, s) \in Q_m\} \subseteq L({\bf G})$.
Let $K \subseteq \Sigma^*$ be a language; its {\it prefix
closure} is $\overline{K} := \{s\in\Sigma^*|
(\exists t\in\Sigma^*)~st \in K\}$. $K$ is said to be $L_m({\bf G})$-{\it closed} if
$\overline{K} \cap L_m({\bf G}) = K$.
TDES $\bf G$ is \emph{nonblocking} if $\overline{L_m({\bf G})} = L({\bf G})$.

A TDES $\bf G$ can be graphically represented by both its {\it activity
transition graph} (ATG), namely the ordinary transition graph of ${\bf G}_{act}$, and its
{\it timed transition graph} (TTG), namely the ordinary transition graph
of $\bf G$, incorporating the $tick$ transition explicitly.

For two TDES ${\bf G}_1$ and ${\bf G}_2$ with ATG ${\bf G}_{1,act}$
and ${\bf G}_{2,act}$ defined on $\Sigma_{1,act}$ and
$\Sigma_{2,act}$ respectively, their {\it composition} ${\bf Comp}({\bf G}_1,
{\bf G}_2)$, is a new TDES ${\bf G}$ such that ${\bf G}_{act} = {\bf G}_{1,act} || {\bf G}_{2,act}$,
where ``$||$" denotes the synchronous product of two generators \cite{Wonham16a}.
The time bounds on the events of $\bf G$ are determined by: if $\sigma \in \Sigma_{1,act}
\cap\Sigma_{2,act}$, then ${l_{{\bf G},\sigma}} = max(l_{{\bf G}_1,\sigma},{l_{{\bf G}_2,\sigma}})$ and ${u_{{\bf G},\sigma}} =
min({u_{{\bf G}_1,\sigma}}, {u_{{\bf G}_2,\sigma}})$; if $\sigma \in \Sigma_{1,act}\setminus \Sigma_{2,act}$,
then ${l_{{\bf G},\sigma}} = {l_{{\bf G}_1,\sigma}}$ and ${u_{{\bf G},\sigma}} = {u_{{\bf G}_1,\sigma}}$;
if $\sigma \in \Sigma_{2,act}\setminus \Sigma_{1,act}$,
then ${l_{{\bf G},\sigma}} = {l_{{\bf G}_2,\sigma}}$ and ${u_{{\bf G},\sigma}} = {u_{{\bf G}_2,\sigma}}$.
If this leads to ${l_{{\bf G},\sigma}} > {u_{{\bf G},\sigma}}$, the composition ${\bf G}$
does not exist.\footnote{We stress that ${\bf Comp}({\bf G}_1, {\bf G}_2)$ is in general different
from the result of ${\bf G}_1 || {\bf G}_2$, for the latter would force the synchronization
of $tick$ transition as it occurs in the components. Specifically, when $\Sigma_{1,act} \cap
{\Sigma_{2,act}} = \emptyset$, ${\bf Comp}({\bf G}_1, {\bf G}_2) \approx {\bf G}_1 || {\bf G}_2$
where $\approx$ denotes that the closed and marked behavior of the TDES coincide \cite{Wonham16a}.}
Composition of more than two TDES can be similarly
constructed.\footnote{There also exist {\it generalized} TDES (as defined in \cite[Section 9.11]{Wonham16a}), which are represented by only TTG including $tick$ in the alphabet. Namely, a generalized TDES does not have a corresponding ATG or timer information, and is simply an ordinary finite-state generator whose event set includes $tick$. Generalized TDES are often adopted to model temporal specifications and supervisors, and represent controlled plant behaviors. To compose two or more generalized TDES, we use the synchronous product ``$||$", rather than $\bf Comp$. }

To use TDES $\bf G$ in (\ref{eq:G}) for supervisory control, first
designate a subset of events, denoted by $\Sigma_{hib} \subseteq \Sigma_{act}$,
to be the {\it prohibitible} events which can be disabled by an external supervisor.
Next, and specific to TDES, specify a subset of {\it forcible} events,
denoted by $\Sigma_{for} \subseteq \Sigma_{act}$, which can {\it preempt}
the occurrence of event $tick$.
Now it is convenient to define the {\it controllable} event set $\Sigma_c :=
\Sigma_{hib}~\dot\cup~\{tick\}$. 
The {\it uncontrollable} event set is $\Sigma_{uc} := \Sigma \setminus \Sigma_c$.
A sublanguage $K \subseteq L_m({\bf G})$ is {\it controllable} if, for
all $s \in \overline{K}$,
\begin{eqnarray*} 
Elig_K(s)\supseteq
\left\{
   \begin{array}{lcl}
      Elig_{\bf G}(s)\cap(\Sigma_{uc} \dot{\cup}\{tick\}) \\
      ~~~~~~~~~~~~~~~~~\text{if} ~~Elig_K(s)\cap \Sigma_{for} = \emptyset,\\
      Elig_{\bf G}(s)\cap\Sigma_{uc}               \\
      ~~~~~~~~~~~~~~~~~\text{if} ~~Elig_K(s)\cap \Sigma_{for} \neq \emptyset,
   \end{array}
\right.
\end{eqnarray*}
where $Elig_K(s):= \{\sigma \in \Sigma|s\sigma \in \overline{K}\}$
is the subset of eligible events after string $s$.

For partial observation, $\Sigma$ is partitioned into $\Sigma_o$, the subset
of observable events, and $\Sigma_{uo}$, the subset of unobservable events (i.e.
$\Sigma = \Sigma_o \dot\cup \Sigma_{uo}$). Bring in the \emph{natural projection}
$P : \Sigma^* \rightarrow \Sigma_o^*$ defined by: (i) $P(\epsilon) = \epsilon$;
(ii) $P(\sigma) = \sigma$ if $\sigma \in \Sigma_o$ and otherwise $P(\sigma) = \epsilon$;
(iii) for all $s\in \Sigma^*$ and $\sigma \in \Sigma$, $P(s\sigma) = P(s)P(\sigma)$.
As usual, $P$ is extended to $P : Pwr(\Sigma^*) \rightarrow Pwr(\Sigma_o^*)$, where
$Pwr(\cdot)$ denotes powerset. Write $P^{-1}: Pwr(\Sigma_o^*) \rightarrow Pwr(\Sigma^*)$
for the \emph{inverse-image function} of $P$.
A language $K \subseteq L_m({\bf G})$ is {\it observable} if for every pair of strings
$s, s' \in \Sigma^*$ with $P(s) = P(s')$ there holds
\begin{equation*} 
(\forall \sigma \in {\Sigma_{act}\cup\{tick\}}) s\sigma \in \overline{K}, s' \in \overline{K},
s'\sigma \in L({\bf G}) \Rightarrow s'\sigma \in \overline{K}
\end{equation*}
where $P:\Sigma^*\rightarrow \Sigma_o^*$ is the corresponding natural projection.

A {\it supervisor} $V$ under partial observation is any map
$V:P(L({\bf G}))\rightarrow Pwr(\Sigma)$. Then the closed-loop system is
$V/{\bf G}$ with closed behavior $L(V/{\bf G})$ and marked
behavior $L_m(V/{\bf G})$ ($:=L(V/{\bf G})\cap L_m({\bf G})$) \cite{LinWon95}.
%
%
A supervisor $V$ is {\it nonblocking} if $\overline{L_m(V/{\bf G})} = L(V/{\bf G})$,
and {\it admissible} if for each $s \in L(V/{\bf G})$, $(\textnormal{i})\ \Sigma_{uc} \subseteq V(P(s))$ and
\begin{align*}
(\textnormal{ii}) Elig_{\bf G}(s) \cap V(P(s)) \cap \Sigma_{for} = \emptyset&,\ tick \in Elig_{\bf G}(s) \\
& \Rightarrow tick \in V(P(s)).
\end{align*}

It has been proved \cite{LinWon95} that a nonblocking,
admissible supervisory control $V$ exists which synthesizes a (nonempty)
sublanguage $K\subseteq L_m({\bf G})$ such that $L_m(V/{\bf G}) = K$ if and
only if $K$ is (timed) observable, controllable and $L_m({\bf G})$-closed. While
controllability and $L_m({\bf G})$-closedness are properties closed under
set union, observability is not; consequently when $K$ is not observable,
there generally does not exist the supremal observable (controllable and
$L_m({\bf G})$-closed) sublanguage of $K$.

Recently in \cite{CaiZW16}, we proposed a new concept of
{\it timed relative observability}, which is stronger
than  timed observability, but permits the existence of the supremal relatively
observable sublanguage.
Let $C \subseteq L_m({\bf G})$. A language $K \subseteq C$ is {\it timed relatively
observable} (or timed $C$-observable), if for every pair of strings
$s, s' \in \Sigma^*$ with $P(s) = P(s')$ there holds 
\begin{align} \label{eq:def_relobs}
(\forall \sigma \in \Sigma_{act}\cup & \{tick\}) \notag \\
&s\sigma \in \overline{K}, s' \in \overline{C},
s'\sigma \in L({\bf G}) \Rightarrow s'\sigma \in \overline{K}.
\end{align}
In this paper, only timed relative observability (or timed $C$-observability) is used; thus for simplicity we shall
henceforth often omit the word ``timed".

For an arbitrary sublanguage $E \subseteq L_m({\bf G})$, write $\mathcal{CO}(E)$ for the family
of $C$-observable, controllable and $L_m({\bf G})$-closed sublanguages of $E$.
Then $\mathcal{CO}(E)$
is nonempty (the empty language $\emptyset$ belongs) and is closed under set union;
$\mathcal{CO}(E)$ has a unique supremal element $\sup \mathcal{CO}(E)$
given by
\[\sup\mathcal{CO}(E) = \bigcup\{K|K\in\mathcal{CO}(E)\}\]
which may be effectively computed \cite{CaiZW15,CaiZW16}.
Note that since relative observability is stronger than observability,
$\sup \mathcal{CO}(E)$ is observable (controllable and $L_m({\bf G})$-closed),
and since relative observability is weaker than normality, $\sup \mathcal{CO}(E)$
is generally larger than its normality counterpart.



\section{Formulation of Partial-Observation Supervisor Localization Problem} \label{sec:ProbFormu}


Let the plant $\textbf{G}$ be comprised of $N$ component TDES
\begin{equation}\label{eq:multi_agent}
{\bf G}_k = (Q_k, \Sigma_k, \delta_k, q_{0,k}, Q_{m,k}),\ \ \ k =
1,...,N.
\end{equation}
Then ${\bf G} = {\bf Comp} ({\bf G}_1, ..., {\bf G}_N)$, where $\bf Comp$
is the composition operator defined in Section \ref{sec:prelim}
which is used to build complex TDES from simpler ones. Let $\Sigma_o \subseteq \Sigma (:= \Sigma_1 \cup ... \cup \Sigma_N)$
be the subset of observable events and $P: \Sigma^*\rightarrow \Sigma_o^*$
the corresponding natural projection. 
Note that $\Sigma_k$ are not pairwise disjoint, because event $tick$ is shared by all components ${\bf G}_k$
(each TTG ${\bf G}_k$ is constructed from its ATG ${\bf G}_{k,act}$ and
the corresponding time bounds by the rules in \cite{BraWon94,Wonham16a}
and thus contains event $tick$); and $tick$ may or may not be observable.

These components are implicitly coupled through a specification language $E \subseteq \Sigma^*$
that imposes a constraint on the global behavior of $\textbf{G}$ ($E$ may itself be the
composition of multiple component specifications). For the plant $\textbf{G}$ and the imposed
specification $E$, let the generator $\textbf{SUP} = (X, \Sigma, \xi, x_0, X_m)$ be such that
\begin{equation} \label{eq:monosup}
L_m(\textbf{SUP}) := \sup \mathcal {CO}(E \cap L_m(\textbf{G})). \footnotemark
\end{equation}
\footnotetext{${\bf SUP}$ can be computed by an algorithm presented in \cite{CaiZW16} (Algorithm 2), with the
ambient language $C$ set to be the supremal controllable and $L_m({\bf G})$-closed sublanguage of $E\cap L_m({\bf G})$.}
We call $\bf SUP$ the {\it controllable and observable behavior}.
Note that $\bf SUP$ is not a `partial-observation supervisor' (to be defined in the next section),
which can only contain observable events as state changers. To rule out the
trivial case, we assume that $L_m(\textbf{SUP}) \neq \emptyset$.

The control actions of {\bf SUP} include (i) disabling prohibitible
events in $\Sigma_{hib}$ and (ii) preempting event $tick$ via forcible events in
$\Sigma_{for}$. Accordingly, the localization of $\bf SUP$'s
control actions under partial-observation is with respect to not only each prohibitible
event's disabling action (just as the untimed counterpart in \cite{ZhangCW17}), but also each forcible event's preemptive action.
The latter is specific to TDES, for which we introduce below
the new concept of ``partial-observation local preemptor".

Let $\alpha \in \Sigma_{for}$ be an arbitrary forcible event,
which may or may not be observable. We say that a generator
\[{\bf LOC}_\alpha^P = (Y_\alpha,\Sigma_\alpha,\eta_\alpha,y_{0,\alpha},Y_{m,\alpha}),\
\Sigma_\alpha \subseteq \Sigma_o \cup \{\alpha, tick\}\]
is a {\it partial-observation local preemptor} for $\alpha$ if (i) ${\bf LOC}_\alpha^P$
preempts event $tick$ consistently with $\bf SUP$ when
$tick$ is preempted by $\alpha$, and (ii) if $\sigma \in \{\alpha, tick\}$
is unobservable,
then $\sigma$-transitions can only be selfloops in ${\bf LOC}_\alpha^P$
(other unobservable events in $\Sigma\setminus \Sigma_\alpha$ are not
defined in, and thus not selfloops in, ${\bf LOC}_\alpha^{P}$).

First, condition (i) means that for all $s \in \Sigma^*$ if $s\alpha \in L({\bf SUP})$,
there holds
{\small \begin{equation} \label{eq:Tloc1}
P_\alpha(s).tick \in L({{\bf LOC}_\alpha^P}), s.tick \in L({\bf G}) \Leftrightarrow s.tick \in L({\bf SUP})
\end{equation}}
where $P_\alpha:\Sigma^*\rightarrow \Sigma_\alpha^*$ is the natural projection.
Notation $s.tick$ means that event $tick$ occurs after string $s$ and will be
used henceforth. Note that specific to TDES, only when $s\alpha \in L({\bf SUP})$
can $tick$-occurrence after $s$ be preempted by $\alpha$ in ${\bf LOC}_\alpha^P$.
Also note that ${\bf LOC}_{\alpha}^P$ is not required to
preempt $tick$ consistently with $\bf SUP$ when $tick$ is preempted by other forcible event
$\alpha'$; thus ${\bf LOC}_{\alpha}^P$ is only responsible for the preemption of $tick$ by
$\alpha$.
Second, condition (ii) requires that
only observable events may cause state change in ${\bf LOC}_\alpha^P$, i.e.
\begin{equation} \label{eq:Tloc2}
(\forall y, y' \in Y_\alpha, \forall \sigma \in \Sigma_\alpha)\ y' = \eta_\alpha(y,\sigma)!,
y \neq y' \Rightarrow \sigma \in \Sigma_o.
\end{equation}
This requirement is a distinguishing feature of
a partial-observation local preemptor as compared to its
full-observation counterpart in \cite{ZhangEt13}.

Note that the event set $\Sigma_\alpha$ of ${\bf LOC}_\alpha^P$ in
general satisfies
\[\{\alpha, tick\} \subseteq \Sigma_\alpha \subseteq \Sigma_o \cup
\{\alpha, tick\};\]
in typical cases, both subset containments are strict. In fact, the
events in $\Sigma_\alpha \setminus \{ \alpha, tick\}$ are communication
events that may be critical to achieve synchronization with other
partial-observation local preemptors/controllers. The $\Sigma_\alpha$ is
not fixed \emph{a priori}, but will be determined as part of the
localization result presented in the next section.

Next, let $\beta \in \Sigma_{hib}$ be an arbitrary prohibitible event,
which may or may not be observable. A generator
\[{\bf LOC}_\beta^C =
(Y_\beta,\Sigma_\beta,\eta_\beta,y_{0,\beta},Y_{m,\beta}),\
\Sigma_\beta \subseteq \Sigma_o \cup \{\beta\}\]
is a {\it partial-observation local controller} for $\beta$ if (i)
${\bf LOC}_\beta^C$ enables/disables the event $\beta$ (and only
$\beta$) consistently with $\bf SUP$, and (ii) if $\beta$ is unobservable,
then $\beta$-transitions can only be selfloops in ${\bf LOC}_\beta^C$.
Here condition (i) means that for all $s \in \Sigma^*$
there holds
\begin{equation} \label{eq:Cloc1}
P_\beta(s)\beta \in L({{\bf LOC}_\beta^C}), s\beta \in L({\bf G}) \Leftrightarrow s\beta \in L({\bf SUP})
\end{equation}
where $P_\beta:\Sigma^*\rightarrow \Sigma_\beta^*$ is the natural projection.
Note that ${\bf LOC}_{\beta}^C$ is not required to
disable/enable other prohibitible event $\beta'$ consistently with $\bf SUP$;
thus ${\bf LOC}_{\beta}^C$ is only responsible for the disablement/enablement
of $\beta$.
Condition (ii) imposes the same requirement (ii) of ${\bf LOC}_\alpha^P$ on ${\bf LOC}_\beta^C$,
i.e. equation (\ref{eq:Tloc2}) holds for all $y,y'\in Y_\beta$ and $\sigma \in \Sigma_\beta$.

The event set $\Sigma_\beta$ of ${\bf LOC}_\beta^C$ in
general satisfies
$\{\beta\} \subseteq \Sigma_\beta \subseteq \Sigma_o \cup \{\beta\};$
in typical cases, both subset containments are strict. Like $\Sigma_\alpha$
above, $\Sigma_\beta$ will be generated as part of our localization result.

The definition of partial-observation local controller differs from that of partial-observation
local preemptor in condition (i) (conditions (ii) are identical because they are required for partial
observation). Condition (i) of partial-observation local preemptor specially
requires that the consistency on $tick$ preemption is considered only when a forcible event $\alpha$ is enabled.
Since every forcible event may preempt $tick$, there will exist a set of partial-observation local preemptors responsible
for preempting the event $tick$, one for each relevant forcible event. While for any prohibitible event in $\Sigma_{hib}$, there is only
one partial-observation local controller responsible for disabling/enabling it.


We are now ready to formulate the {\it Partial-Observation Supervisor Localization
Problem}:

{\it Construct a set of partial-observation local preemptors
$\{{\bf LOC}_\alpha^P|\alpha \in \Sigma_{for}\}$ and a set of partial-observation
local controllers $\{{\bf LOC}_\beta^C |\ \beta \in \Sigma_{hib}\}$
with
\begin{align}
L({\bf LOC}):= &\Big(\mathop \bigcap\limits_{\alpha \in \Sigma_{for}}P_\alpha^{-1}L({\bf LOC}^P_{\alpha}) \Big)\notag\\
          \cap &\Big(\mathop \bigcap\limits_{\beta \in \Sigma_{hib}}P_\beta^{-1}L({\bf LOC}^C_{\beta}) \Big) \label{eq:sub1:loc}\\
L_m({\bf LOC}):= &\Big(\mathop \bigcap\limits_{\alpha \in \Sigma_{for}}P_\alpha^{-1}L_m({\bf LOC}^P_{\alpha})\Big)\notag\\
          \cap &\Big(\mathop \bigcap\limits_{\beta \in \Sigma_{hib}}P_\beta^{-1}L_m({\bf LOC}^C_{\beta}) \Big) \label{eq:sub2:loc}
\end{align}
such that the collective controlled behavior of $\bf LOC$ is equivalent to the controllable
and observable controlled behavior $\bf SUP$ in (\ref{eq:monosup}) with respect to $\bf G$, i.e.
\begin{align*}
   L({\bf G}) \cap L({\bf LOC}) &= L({\bf SUP}), \\
   L_m({\bf G}) \cap L_m({\bf LOC}) &= L_m({\bf SUP}).
\end{align*}
}

\begin{figure}[!t]
\begin{center}
\includegraphics[scale = 0.75]{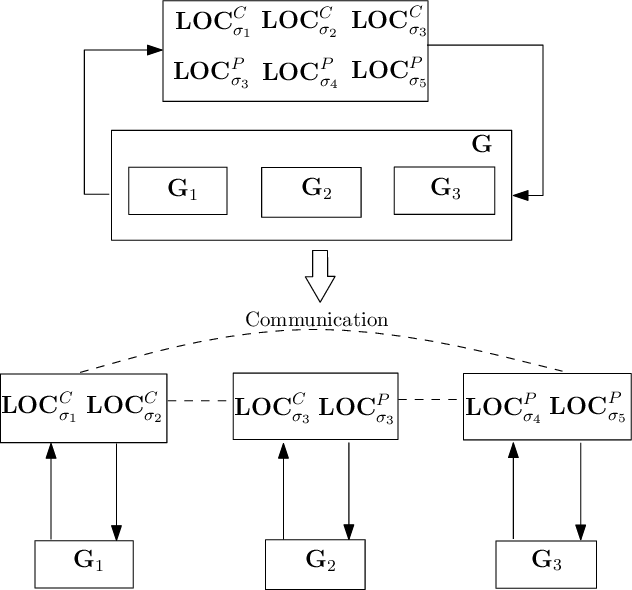}
\caption{Example of distributed control by allocating local
preemptors/controllers. Let plant $\textbf{G}$ be composed of
three components ${\bf G}_k$ with event sets $\Sigma_k$, $k \in [1,3]$.
Suppose $\sigma_1,\sigma_2 \in \Sigma_{hib,1}$, $\sigma_2, \sigma_3 \in
\Sigma_{hib,2}$, $\sigma_3 \in \Sigma_{for,2}$ and $\sigma_3, \sigma_4, \sigma_5 \in \Sigma_{for,3}$; thus
${\bf G}_1$ and ${\bf G}_2$ share event $\sigma_2$, and ${\bf G}_2$
and ${\bf G}_3$ share event $\sigma_3$ (event $tick$ is shared
by all components).
Then a convenient allocation as in (\ref{eq:disforce})
is displayed, where $\sigma_1,\sigma_2 \in \hat\Sigma_{hib,1}$,
$\sigma_3 \in \hat\Sigma_{hib,2}$, $\sigma_3 \in \hat\Sigma_{for,2}$
and $\sigma_4, \sigma_5 \in \hat\Sigma_{for,3}$, and
each local controller/preemptor is owned by
exactly one component.
} \label{fig:allocation}
\end{center}
\end{figure}

Having a set of partial-observation local preemptors $\{{\bf LOC}_\alpha^P|\alpha \in \Sigma_{for}\}$,
and a set of partial-observation local controllers $\{{\bf LOC}_\beta^C |\ \beta \in \Sigma_{hib}\}$,
we build for the TDES plant ${\bf G}$ (as in (\ref{eq:multi_agent})) with multiple components
${\bf G}_k$ ($k = 1,...,N$) a nonblocking distributed control architecture under partial
observation. Let $\Sigma_{for,k} = \Sigma_k \cap \Sigma_{for}$ and
$\Sigma_{hib,k} = \Sigma_k \cap \Sigma_{hib}$ be the subset of forcible events and subset of
prohibitible events of ${\bf G}_k$, respectively. One way of allocating the local preemptors/controllers to the components
is as follows. First, construct
a set of disjoint subsets of forcible events $\{\hat\Sigma_{for,k}|k = 1,..., N\}$
according to:
\begin{align} \label{eq:disforce}
\hat\Sigma_{for,1} &:= \Sigma_{for,1}; \notag\\
\hat\Sigma_{for,2} &:= \Sigma_{for,2}\setminus \hat\Sigma_{for,1};\notag\\
&\vdots \\
\hat\Sigma_{for,N} &:= \Sigma_{for,N}\setminus \big(\hat\Sigma_{for,1} \cup \hat\Sigma_{for,2} \cup ... \cup \hat\Sigma_{for,N-1}\big) \notag
\end{align}
Similarly, a set of disjoint subsets of prohibitible events $\{\hat\Sigma_{hib,k}|k = 1,..., N\}$
can be constructed.
Second, let each local preemptor (resp. controller) belong to the component ${\bf G}_k$ such that
$\hat\Sigma_{for,k}$ (resp. $\hat\Sigma_{hib,k}$) contains the corresponding forcible
(resp. prohibitible) event; an example is displayed in Fig.~\ref{fig:allocation}.
By this allocation policy, each local preemptor/controller will
be owned by exactly one component, thereby we build a distributed control architecture for $\bf G$.
Note that different orders of choosing $\hat\Sigma_{for,k}$ and $\hat\Sigma_{hib,k}$
generally lead to different allocation policies, the choice of which is case-dependent.
We shall use this allocation rule in the example (Timed Workcell) below.

\section{Partial-Observation Localization Procedure} \label{sec:LocProc}


We solve the Partial-Observation Supervisor Localization Problem of TDES
by developing a partial-observation localization procedure for the preemptive
and disabling action, respectively. The procedure extends the untimed
counterpart in \cite{ZhangCW17}. In particular, localizing the
preemption of event $tick$ with respect to each forcible event under partial
observation is novel in the current TDES setup, for which we introduce below
the concept of ``partial-observation preemption cover''.

Let ${\bf G} = (Q,\Sigma,\delta,q_0,Q_m)$ be the TDES plant,
$\Sigma_o \subseteq \Sigma$ the subset of observable events, and $P:
\Sigma^* \rightarrow \Sigma^*_o$ the corresponding natural
projection. Also let ${\bf SUP} = (X,\Sigma,\xi,x_0,X_m)$ be controllable
 and observable behavior (as defined in (\ref{eq:monosup})).
We present the localization of preemptive and disabling action in the sequel.
To this end, we need the concept of {\it uncertainty set}.

For $s \in L({\bf SUP})$, let $U(s)$ be the subset of
states of $\bf SUP$ that may be reached by some string $s'$ that looks like $s$,
i.e.
\[U(s) = \{x\in X|(\exists s' \in \Sigma^*) P(s) = P(s'), x =
\xi(x_0,s')\}.\]
We call $U(s)$ the {\it uncertainty set} \cite{ZhangCW17} of the state
$\xi(x_0,s)$ associated with string $s$. Let
$\mathcal{U}(X) := \{U(s) \subseteq X | s\in L({\bf SUP})\}, $
i.e. $\mathcal{U}(X)$ is the set of uncertainty sets of all states (associated
with strings in $L({\bf SUP})$)
in $X$. The size of $\mathcal{U}(X)$ is in general $|\mathcal{U}(X)| \leq
2^{|X|}$.

The transition function associated with $\mathcal{U}(X)$ is
$\hat\xi:\mathcal{U}(X)\times \Sigma_o \rightarrow \mathcal{U}(X)$
given by
\begin{equation} \label{eq:Utransit}
\hat\xi(U,\sigma) = \bigcup \{\xi(x,u_1 \sigma u_2) | x\in U,
u_1,u_2\in \Sigma_{uo}^*\}.
\end{equation}
With $\mathcal{U}(X)$ and $\hat\xi$, define the {\it partial-observation
monolithic supervisor} \cite{Wonham16a,CasLaf08}
\begin{align} \label{eq:posup}
{\bf SUPO} = (\mathcal{U}(X),\Sigma_o,\hat\xi,U_0,U_m),
\end{align}
where $U_0 = U(\epsilon)$ and $U_m = \{U\in \mathcal{U}(X)|U\cap X_m\neq \emptyset\}$.
${\bf SUPO}$ can be constructed by the well-known subset construction algorithm in \cite{Hopcroft14}
and it is known \cite{Wonham16a,CasLaf08} that $L({\bf SUPO}) = P(L({\bf SUP}))$
and $L_m({\bf SUPO}) = P(L_m({\bf SUP}))$.

Now let $U\in\mathcal{U}(X)$, $x \in U$ be any state in $\bf SUP$
and $\sigma \in \Sigma_c $ ($= \Sigma_{hib}
\dot\cup \{tick\}$) be a controllable event.
We say that
\begin{enumerate}[(i)]

\item $\sigma$ is {\it enabled} at $x \in U$ if $\sigma$ is defined at
$x$ in $\bf SUP$;

\item $\sigma$ ($\neq tick$) is
{\it disabled} at $x \in U$ if it is not defined at $x$ in $\bf SUP$,
but is defined at some state $q$ in $\bf G$ that {\it corresponds} to $x \in U$
(i.e. there exists a string $s \in \Sigma^*$ such that $\xi(x_0,s) = x$,
and $\delta(q_0,s) = q$);

\item $\sigma$ is {\it not defined} at $x \in U$
if it is not defined at $x$ in $\bf SUP$, and also
not defined at any state in $\bf G$ that corresponds to $x$;

\item $\sigma = tick$  is
{\it preempted} at $x \in U$ if $tick$ is not defined at $x$ in $\bf SUP$,
but is defined at some state $q$ in $\bf G$ that corresponds to $x$, and additionally there must exist a forcible event
$\sigma_f$ that is defined at $x$ in $\bf SUP$.

\end{enumerate}

The formal definitions of (i)-(iii) can be found in \cite{ZhangCW17}.
Since (iv) is specific to TDES (under partial observation), we define it as follows:
$\sigma$ ($ = tick$) is {\it preempted} at $x \in U$
if $ \neg \xi(x,tick)!$ and
\begin{align*}
(\exists s \in \Sigma^*) (\exists \sigma_f \in
\Sigma_{for}) \xi(x_0,s)=x  ~\&~\hat\xi(U_0, Ps) = U\\
~\&~ \xi(x,\sigma_f)!\ \&\ \delta(q_0,s.tick)!.
\end{align*}

\begin{lemma} \label{lem:Tproperty}
Given ${\bf SUP}$ in (\ref{eq:monosup}), let $U \in \mathcal{U}(X)$, $x \in U$,
and $\sigma \in \Sigma_c$. If $\sigma$ is enabled at $x\in U$, then for all $x' \in U$,
either $\sigma$ is also enabled at $x'\in U$, or $\sigma$ is not defined at $x' \in U$.
On the other hand, if $\sigma$ is disabled (resp. preempted) at $x \in U$, then for
all $x' \in U$, either $\sigma$ is also disabled (resp. preempted) at $x' \in U$,
or $\sigma$ is not defined $x' \in U$.
\end{lemma}

{\it Proof.} We prove the first statement; the second can be proved by
a similar argument.

By $x \in U \in \mathcal{U}(X)$, there exists $s\in L({\bf SUP})$ such that $\xi(x_0,s) = x$ and $U(s) = U$.
Suppose that  $\sigma \in \Sigma_c$ is enabled at $x \in U$, i.e. $\xi(x,\sigma)!$;
it follows that $\xi(x_0,s\sigma)!$, i.e. $s\sigma \in L({\bf SUP})$.
Now let $x'$ be an arbitrary state in $U = U(s)$. According to the subset construction algorithm, there must exist
$s' \in L({\bf SUP})$ such that $\xi(x_0,s') = x'$ (i.e. $s' \in L({\bf SUP})$) and $Ps' = Ps$.
At state $x'$, either (i) $\xi(x',\sigma)!$, or (ii) $\neg \xi(x',\sigma)!$.
Case (i) means that $\sigma$ is enabled at $x' \in U$. In case (ii), we claim that
$s'\sigma \notin L({\bf G})$, i.e. $\sigma$ is not defined at $x' \in U$.
To see this, assume on the contrary that $s'\sigma \in L({\bf G})$. Then we have $Ps'=Ps$, $s' \in L({\bf SUP})$, $s'\sigma \in L({\bf G})$, $s' \sigma \notin L({\bf SUP})$, and $s\sigma \in L({\bf SUP})$. This implies that $L_m({\bf SUP})$ is not observable, which is a contradiction to the definition of $L_m({\bf SUP})$ in (5).
Therefore, in case (ii), $\sigma$ is not defined at $x' \in U$ after all.
\hfill $\square$

%


\subsection{Partial-Observation Localization of Preemptive Action} \label{subsec:PreemptAct}


Under partial observation, the preemptive action after string $s \in L({\bf SUP})$
depends not on the single state $\xi(x_0,s)$, but on the uncertainty set $U(s)$.

Fix an arbitrary forcible event $\alpha \in \Sigma_{for}$. First define
$E_{tick}: \mathcal{U}(X) \rightarrow \{0,1\}$ according to
\begin{equation*} \label{eq:ET1}
\begin{split}
(\forall U \in \mathcal{U}(X))~
E_{tick}(U) &= \left\{
  \begin{array}{ll}
    1, & \hbox{if $(\exists x \in U) \xi(x,tick)!$,} \\
    0, & \hbox{otherwise.}
  \end{array}
\right.\\
\end{split}
\end{equation*}
Thus $E_{tick}(U) = 1$ means that $tick$ is enabled at some state $x \in U$, i.e.
$tick$ is eligible to occur and its occurrence will not be preempted by
any forcible events. Then
by Lemma~\ref{lem:Tproperty}, at any other state $x' \in U$, $tick$ is either
enabled or not defined. Then define $F_{\alpha}: \mathcal{U}(X) \rightarrow \{0,1\}$
according to
\begin{equation*} \label{eq:ET2}
\begin{split}
&(\forall U \in \mathcal{U}(X)) \\
&F_\alpha(U)= \left\{
  \begin{array}{ll}
    1, & \hbox{if $(\exists x \in U) ~\xi(x,\alpha)!~\&~\neg\xi(x,tick)! \ \& $}\\
     & ( (\exists s\in \Sigma^*)\xi(x_0,s)= x \ \&~\hat\xi(U_0, Ps) = U \\
     & ~~~~~~~~~~~~~~~~~~~~~~~~~~~~~~\&~\delta(q_0,s.tick)!) \\
    0, & \hbox{otherwise.}
  \end{array}
\right.\\
\end{split}
\end{equation*}
Hence $F_\alpha(U) = 1$ means that $tick$ is preempted by the occurrence of $\alpha$ at some
state $x \in U$, i.e. there exists a state $x\in U$ such that
$tick$ is eligible to occur at some state in $\bf G$ that corresponds to $x$,
but its occurrence is effectively preempted by $\alpha$ that has already been enabled at $x$. Again by Lemma~\ref{lem:Tproperty},
at any other state $x'\in U$, $tick$ is either preempted or not defined.
Note that at state $x$, $\alpha$ need not be the only forcible event that
preempts $tick$, for there can be other forcible events, say $\alpha'$, defined at $x$.
In that case, $F_{\alpha'}(U) = 1$ holds as well.

Based on the preemption information captured by $E_{tick}$ and $F_\alpha$ above,
we define the {preemption consistency relation} $\mathcal{R}_\alpha^P \subseteq
\mathcal{U}(X)\times \mathcal{U}(X)$ (for $\alpha$) as follows.

\begin{definition} \label{def:PreemptConsist}
For $U,U' \in \mathcal{U}(X)$, we say that $U$ and $U'$ are {\it preemption consistent}
with respect to $\alpha$, written $(U,U')\in \mathcal{R}_\alpha^P$, if
\begin{align*}
E_{tick}(U)\cdot F_\alpha(U') = 0 = E_{tick}(U')\cdot F_\alpha(U).
\end{align*}
\end{definition}
Thus a pair of uncertainty sets $(U,U')$ satisfies $(U,U')\in \mathcal{R}_\alpha^P$
if $tick$ is defined at some state of $U$, but not preempted by $\alpha$ at any
state of $U'$, and vice versa. It is easily verified that $\mathcal{R}_\alpha^P$ is
reflexive and symmetric, but not transitive. Hence $\mathcal{R}_\alpha^P$ is not an
equivalence relation. This fact leads to the definition of a {\it partial-observation
preemption cover}. Recall that a {\it cover} on a set $\mathcal{U}(X)$ is a family of
nonempty subsets (or {\it cells}) $\mathcal{U}_{i}$ ($i \in I_\alpha$, $I_\alpha$ is an index set)
of $\mathcal{U}(X)$ whose union is $\mathcal{U}(X)$, i.e.
$\mathcal{U}(X) = \bigcup\{\mathcal{U}_{i} | \mathcal{U}_{i} \subseteq\mathcal{U}(X), \mathcal{U}_{i} \neq \emptyset, {i}\in I_\alpha\}$.


\begin{definition} \label{def:controlcover}
Let $I_\alpha$ be some index set, and $\mathcal {C}_\alpha^P =
\{\mathcal{U}_{i} \subseteq\mathcal{U}(X) | {i}\in I_\alpha\}$ be a cover on
$\mathcal{U}(X)$. We say that $\mathcal {C}_\alpha^P$ is a {\it
partial-observation preemption cover} with respect to $\alpha$ if
\begin{align*}
(\textnormal{i})~~ & (\forall {i} \in I_\alpha, \forall U,U' \in \mathcal{U}_{i})~ (U,U') \in \mathcal{R}_\alpha^P,\\
(\textnormal{ii})~~ & (\forall {i} \in I_\alpha, \forall \sigma \in \Sigma_o) (\exists U\in
\mathcal{U}_{i}) ~\hat\xi(U,\sigma)\neq\emptyset\Rightarrow\\
 &\big((\exists j \in I_\alpha)(\forall U'\in \mathcal{U}_{i})~\hat\xi(U',\sigma)\neq\emptyset
                  \Rightarrow \hat\xi(U',\sigma) \in \mathcal{U}_{j}\big).
\end{align*}
\end{definition}
A partial-observation preemption cover $\mathcal {C}_\alpha^P$ lumps the
uncertainty sets $U \in \mathcal{U}(X)$ into (possibly overlapping)
{\it cells} $\mathcal{U}_{i} \in \mathcal {C}_\alpha^P$, $i \in I_\alpha$,
according to (i) the uncertainty sets $U$ that reside in the same cell
$\mathcal{U}_{i}$ must be pairwise preemption consistent, and (ii) for every
observable event $\sigma\in \Sigma_o$, the uncertainty sets $U'$ that
can be reached from any uncertainty set $U\in \mathcal{U}_{i}$ by a
one-step transition $\sigma$ must be covered by the same cell
$\mathcal{U}_{j}$. Inductively, two uncertainty sets $U$ and $U'$
belong to a common cell of $\mathcal{C}_\alpha^P$ if and only if $U$
and $U'$ are preemption consistent, and two future uncertainty sets
that can be reached respectively from $U$ and $U'$ by a given observable
string are again preemption consistent.

The partial-observation preemption cover $\mathcal {C}_\alpha^P$ differs
from its full-observation counterpart in \cite{ZhangEt13} in two
aspects. First, $\mathcal {C}_\alpha^P$ is defined on
$\mathcal{U}(X)$, not on $X$; this is due to state uncertainty
caused by partial observation. Second, in condition (ii) of
$\mathcal {C}_\alpha^P$ only observable events in $\Sigma_o$ are
considered, not $\Sigma$; this is to generate partial-observation
local preemptors whose state transitions are triggered only by
observable events.
We call $\mathcal{C}_\alpha^P$ a {\it
partial-observation preemption congruence} if $\mathcal{C}_\alpha^P$
happens to be a partition on $\mathcal{U}(X)$.

Having defined a partial-observation preemption cover $\mathcal{C}_\alpha^P$
on $\mathcal{U}(X)$, we construct a generator ${\bf J}_\alpha = (I_\alpha,
\Sigma_o,\zeta_\alpha,i_{0,\alpha},I_{m,\alpha})$ 
and two functions $\psi_\alpha:I_\alpha\rightarrow \{0,1\}$ and $\psi_{tick}:
I_\alpha\rightarrow \{0,1\}$
as follows. Recall from (\ref{eq:posup}) that $U_0 = U(\epsilon)$ and
thus $x_0 \in U_0$.
\begin{align}
(\textnormal{i})~~ & i_{0,\alpha} \in I_\alpha ~\text{such that}~  {U_0 \in \mathcal{U}_{i_{0,\alpha}}};\label{eq:sub1:construct}\\
(\textnormal{ii})~~ & I_{m,\alpha} := \{i \in I_\alpha |  (\exists U \in \mathcal{U}_{i}) X_m\cap U \neq \emptyset \};\label{eq:sub2:construct}\\
(\textnormal{iiii})~~ & \zeta_\alpha:I_\alpha\times\Sigma_o\rightarrow I_\alpha ~\text{with}~
            \zeta_\alpha(i,\sigma)=j \notag\\
           &\text{if}~ (\exists U \in \mathcal{U}_{i})~\hat{\xi}(U,\sigma)\in \mathcal{U}_{j};
            \label{eq:sub3:construct}\\
(\textnormal{iv})~~ &\psi_\alpha(i) = 1 ~\text{iff}~ (\exists U \in
        \mathcal{U}_{i})(\exists x \in U)~ \xi(x,\alpha)!. \label{eq:sub4:construct}\\
(\textnormal{v})~~ &\psi_{tick}(i) = 1 ~\text{iff}~ (\exists U \in
        \mathcal{U}_{i})~ E_{tick}(U) = 1. \label{eq:sub5:construct}
\end{align}
The function $\psi_\alpha(i) = 1$ means that forcible event $\alpha$
is defined at state $i$ of ${\bf J}_\alpha$,
and  the function $\psi_{tick}(i) = 1$ means that event $tick$
is eligible to occur and its occurrence will not be preempted at
state $i$ of ${\bf J}_\alpha$.
Note that owing to
cell overlapping, the choices of $i_{0,\alpha}$ and $\zeta_\alpha$ may not be
unique, and consequently ${\bf J}_\alpha$ may not be unique. In that
case we simply pick an arbitrary instance of ${\bf J}_\alpha$.

Finally we define the {\it partial-observation local preemptor} ${\bf
LOC}_\alpha^P = (Y_\alpha,\Sigma_\alpha,\eta_\alpha,y_{0,\alpha},Y_{m,\alpha})$
as follows:

\noindent {\bf Step} (i) $Y_\alpha = I_\alpha$, $y_{0,\alpha} = i_{0,\alpha}$, and
$Y_{m,\alpha} = I_{m,\alpha}$. Thus the function $\psi_\alpha$ is $\psi_\alpha:
Y_\alpha\rightarrow\{0,1\}$, and the function $\psi_{tick}$ is $\psi_{tick}:
Y_\alpha\rightarrow\{0,1\}$.

\noindent {\bf Step} (ii) $\Sigma_\alpha = \{\alpha, tick\} \cup \Sigma_{com,\alpha}$, where
\begin{align} \label{eq:comevent_prmpt}
\Sigma_{com,\alpha} := \{\sigma \in \Sigma_o \setminus \{\alpha, tick\} \
|\ (\exists i,j &\in I_\alpha)~ i \neq j \ \&\  \notag\\
  &\zeta_\alpha(i,\sigma)=j\}
\end{align}
Thus $\Sigma_{com,\alpha}$ is the set of observable events that are
not merely selfloops in ${\bf J}_\alpha$ (i.e. these events will cause state changes in
${\bf LOC}_\alpha^P$). It holds by definition
that $\{\alpha, tick\} \subseteq \Sigma_\alpha \subseteq \Sigma_o \cup
\{\alpha,tick\}$, and $\Sigma_{com,\alpha}$ represents the set of {\it communication events}
that need to be communicated to ${\bf LOC}_\alpha^P$.
Note that a communication event $\sigma \in \Sigma_{com,\alpha}$
can be non-forcible or non-prohibitible.

\noindent {\bf Step} (iii) If $\alpha \in \Sigma_o$, then $\eta_\alpha =
\zeta_\alpha|_{Y_\alpha \times \Sigma_\alpha} : Y_\alpha \times
\Sigma_\alpha \rightarrow Y_\alpha$, i.e. $\eta_\alpha$ is the restriction
of $\zeta_\alpha$ to $Y_\alpha \times \Sigma_\alpha$. If $\alpha \in
\Sigma_{uo}$, first obtain $\eta_\alpha = \zeta_\alpha|_{Y_\alpha \times
\Sigma_\alpha}$, then add $\alpha$-selfloops $\eta_\alpha(y,\alpha)=y$
to those $y\in Y_\alpha$ with $\psi_\alpha(y) = 1$.

\noindent {\bf Step} (iv) If $tick \in \Sigma_{uo}$, then add $tick$-selfloops
$\eta_\alpha(y,tick)=y$ to those $y\in Y_\alpha$ with $\psi_{tick}(y) = 1$.

\begin{lemma} \label{lem:preempt}
The generator ${\bf LOC}_\alpha^P$ is a partial-observation local
preemptor for $\alpha$, i.e. (\ref{eq:Tloc1}) and (\ref{eq:Tloc2})
hold.
\end{lemma}

The proof of Lemma~\ref{lem:preempt} will be presented at the end
of this section.

By the same procedure, we generate a set of partial-observation
local preemptors ${\bf LOC}_\alpha^P$, one for each forcible event
$\alpha \in \Sigma_{for}$. We will verify below that these generated
preemptors collectively achieve the same $tick$-preemptive action
as $\bf SUP$ did.


\subsection{Partial-Observation Localization of Disabling Action}\label{subsec:DisableAct}

Next, we turn to the localization of disabling action, which is
analogous to the treatment in \cite{ZhangCW17} for the untimed case. Fix an arbitrary
prohibitible event $\beta \in \Sigma_{hib}$. Define $E_\beta:\mathcal{U}(X)
\rightarrow \{0,1\}$ according to
\[(\forall U\in \mathcal{U}(X))~E_\beta(U) = 1
\mbox{ iff } (\exists x \in U) \xi(x,\beta)!.\]
So $E_\beta(U) = 1$ if event $\beta$
is enabled at some state $x \in U$. Also define $D_\beta:\mathcal{U}(X)
\rightarrow \{0,1\}$ according to
\begin{equation*} \label{eq:ET2}
\begin{split}
(\forall U \in \mathcal{U}(X))& \\
D_\beta(U)=& \left\{
  \begin{array}{ll}
    1, & \hbox{if $(\exists x \in U) \neg\xi(x,\beta)! \ \& $}\\
     & ( (\exists s\in \Sigma^*)\xi(x_0,s)= x ~\&\\
     & ~~~~~~\hat\xi(U_0, Ps) = U \&\ \delta(q_0,s\beta)!) \\
    0, & \hbox{otherwise.}
  \end{array}
\right.\\
\end{split}
\end{equation*}
Hence $D_\beta(U) = 1$ if $\beta$ is disabled at some state $x \in U$.
Now define $M:\mathcal{U}(X) \rightarrow \{0,1\}$ by $M(U) = 1$ iff there
exists $x \in U$ such that $x \in X_m$; and $T:\mathcal{U}(X) \rightarrow \{0,1\}$
by $T(U) = 1$ iff there exists $s \in \Sigma^*$ such that $\xi(x_0, s) \in U$,
$\hat\xi(U_0, Ps) = U$ and $\delta(q_0, s) \in Q_m$.

We define the \emph{control consistency relation}
$\mathcal{R}_\beta^C \subseteq \mathcal{U}(X) \times \mathcal{U}(X)$
with respect to $\beta$ according to
$(U,U')\in \mathcal{R}_\beta^C$ iff
\begin{align*}
&E_\beta(U)\cdot D_\beta(U') = 0 = E_\beta(U')\cdot D_\beta(U)\\
&T(U) = T(U') \Rightarrow M(U) = M(U').
\end{align*}
Let $I_\beta$ be some index set, and $\mathcal {C}_\beta^C =
\{\mathcal{U}_{i} \subseteq\mathcal{U}(X) | i\in I_\beta\}$ a cover on
$\mathcal{U}(X)$. We say that $\mathcal {C}_\beta^C$ is a {\it
partial-observation control cover} with respect to $\beta$ if
\begin{align*}
(\textnormal{i})~ & (\forall {i} \in I_\beta, \forall U,U' \in \mathcal{U}_{i})~ (U,U') \in \mathcal{R}_\beta^C,\\
(\textnormal{ii})~ & (\forall {i} \in I_\beta, \forall \sigma \in \Sigma_o) (\exists U\in
\mathcal{U}_{i}) \hat\xi(U,\sigma)\neq\emptyset\Rightarrow \\
 &~\big( (\exists {j} \in I_\beta)(\forall U'\in \mathcal{U}_{i})\hat\xi(U',\sigma)\neq\emptyset
                  \Rightarrow \hat\xi(U',\sigma) \in \mathcal{U}_{j}\big).
\end{align*}

With the control cover $\mathcal{C}_\beta^C$ on $\mathcal{U}(X)$, we
construct, by the Steps (i)-(iii) above for a local preemptor, a partial-observation local controller
${\bf LOC}_\beta^C = (Y_\beta,\Sigma_\beta,\eta_\beta,y_{0,\beta},Y_{m,\beta})$
for prohibitible event $\beta$. Here, the event set $\Sigma_\beta$
is $\Sigma_\beta = \{\beta\} \cup \Sigma_{com,\beta}$,
where
\begin{align} \label{eq:comevent_disab}
\Sigma_{com,\beta} := \{\sigma \in \Sigma_o \setminus \{\beta\} \
| ~ (\exists i,j \in I_\beta)  i \neq j, 
\zeta_\beta(i,\sigma)=j\}.
\end{align}
It holds by definition that $\{\beta\} \subseteq \Sigma_\beta \subseteq \Sigma_o \cup
\{\beta\}$, and $\Sigma_{com,\beta}$ represents the set of {\it communication events}
that need to be communicated to ${\bf LOC}_\beta^C$.
Similar to the events in $\Sigma_{com,\alpha}$, a communication event $\sigma \in \Sigma_{com,\beta}$
  can be non-forcible or non-prohibitible.

\begin{lemma} \label{lem:loc}
The generator ${\bf LOC}_\beta^C$ is a partial-observation local
controller for prohitibile event $\beta$.
\end{lemma}

For a proof of Lemma~\ref{lem:loc}, see \cite[Lemma 2]{ZhangCW17}.

By the same procedure, we generate a set of partial-observation
local controllers ${\bf LOC}_\beta^C$, one for each prohitibile event
$\beta \in \Sigma_{hib}$. We will verify below that these generated
controllers collectively achieve the same disabling action as $\bf SUP$ did.


\subsection{Main Result} \label{subsec:mainresult}


Here is the main result of this section, which states that
the collective behavior of the partial-observation local preemptors
and local controllers generated by the localization
procedure above is identical to the monolithic controllable and observable $\bf SUP$.

\begin{theorem} \label{thm:equ}
The set of partial-observation local preemptors $\{{\bf
LOC}_\alpha^P|\alpha \in \Sigma_{for}\}$ and the set of partial-observation
local controllers $\{{\bf LOC}_\beta^C|\beta \in \Sigma_{hib}\}$ constructed
above solve the Partial-Observation Supervisor Localization Problem, i.e.
\begin{align}
   L({\bf G}) \cap L({\bf LOC})  &= L({\bf SUP}) \label{eq:sub1:equv}\\
   L_m({\bf G})\cap L_m({\bf LOC}) &= L_m({\bf SUP}) \label{eq:sub2:equv}
\end{align}
where $L({\bf LOC})$ and $L_m({\bf LOC})$ are as defined in (\ref{eq:sub1:loc})
and (\ref{eq:sub2:loc}), respectively.
\end{theorem}


Since for every partial-observation preemption cover (resp. control cover),
the presented procedure constructs a local preemptor (resp.
local controller), Theorem~\ref{thm:equ} asserts that every set of
preemption and control covers together generates a solution to the
Partial-Observation Supervisor Localization Problem.
The localization algorithm in \cite{ZhangCW17} for untimed DES can
easily be adapted in the current TDES case, the only modification being
to use the new definitions of partial-observation preemption and control consistency given in
Sections~\ref{subsec:PreemptAct} and \ref{subsec:DisableAct}. The
complexity of the localization algorithm is $O(n^4)$; since the size $n$ of $\mathcal{U}(X)$
is $n \leq 2^{|X|}$ in general, the algorithm is exponential in $|X|$.

\begin{remark}
As in \cite{CaiWon10a,CaiWon10b,CaiWon16}, for large-scale timed DES in practice,
we may combine our proposed partial-observation supervisor localization with
an efficient decentralized/hierarchical supervisor synthesis approach \cite{FenWon08},
by exploiting modularities that often exist in practical systems and extending the approach in \cite{FenWon08} from untimed
to timed DES. A systematic investigation on this topic is left for our future work.
\end{remark}

Having these obtained partial-observation local preemptors/controllers,
by the allocation policy described in Section III, we build a distributed
control architecture for the multi-component TDES ${\bf G}$ in (\ref{eq:multi_agent}).
As asserted by Theorem~\ref{thm:equ}, the distributed controlled behavior is
identical to the monolithic one, as represented by ${\bf SUP}$.

~

\noindent {\it Proof of Theorem~\ref{thm:equ}}:
First, we prove ($\subseteq$) of (\ref{eq:sub1:equv}), i.e. $L({\bf G}) \cap L({\bf LOC}) \subseteq L({\bf SUP})$, by induction on the length of strings.

For the {\bf base step}, note that none of $L({\bf G})$, $L({\bf LOC})$ and $L({\bf SUP})$ is empty;
and thus the empty string $\epsilon$ belongs to all of them. For the {\bf inductive step},
suppose that $s \in L({\bf G}) \cap L({\bf LOC})$, $s \in L({\bf SUP})$ and
$s\sigma \in L({\bf G}) \cap L({\bf LOC})$ for arbitrary event $\sigma \in \Sigma$;
we must show that $s\sigma \in L({\bf SUP})$. Since $\Sigma = \Sigma_{uc}\dot\cup\Sigma_{hib}\dot\cup\{tick\}$,
$\sigma$ may belong to $\Sigma_{uc}$, $\Sigma_{hib}$ or be equal to $tick$.
The proof for $\sigma \in \Sigma_{uc}$ and $\sigma\in \Sigma_{hib}$ is similar to that
in \cite{ZhangCW17} for untimed DES; in the following, we consider the case $\sigma = tick$, which
is specific to TDES.


By the hypothesis that $s, s.tick \in L({\bf LOC})$,
for every forcible event $\alpha \in \Sigma_{for}$, $s, s.tick \in P_\alpha^{-1}L({\bf LOC}_\alpha^P)$,
i.e. $P_\alpha(s), P_\alpha(s).tick \in L({\bf LOC}_\alpha^P)$. Let $y = \eta_\alpha(y_{0,\alpha},P_\alpha(s))$;
then $\eta_\alpha(y,tick)!$. Since $tick$ may be observable or unobservable,
we consider the following two cases.

(i) $tick \in \Sigma_{uo}$. It follows from the construction rule (iv) of ${\bf LOC}_\alpha^P$ that
$\eta_\alpha(y, tick)!$ implies that for the state $i \in I$ of the generator ${\bf J}_\alpha$
corresponding to $y$ (i.e. $i = \zeta_\alpha(i_0, P(s))$), there holds $\psi_{tick}(i) = 1$.
By the definition of $\psi_{tick}$ in (\ref{eq:sub5:construct}), there exists an uncertainty set
$U \in \mathcal{U}_i$ such that $E_{tick}(U) = 1$. Let $U' = \hat\xi(U_0, Ps)$; by (\ref{eq:sub3:construct})
and $i = \zeta_\alpha(i_0, Ps)$, $U' \in \mathcal{U}_i$. According to (\ref{eq:Utransit}),
$\xi(x_0,s) \in U'$.
Since $U$ and $U'$ belong to the same cell $\mathcal{U}_i$, by the definition of partial-observation
preemption cover they must be preemption consistent, i.e. $(U, U') \in \mathcal{R}_\alpha^P$.
Thus $E_{tick}(U)\cdot F_\alpha(U') = 0$, which implies that $F_\alpha(U') = 0$.
The latter means that for all state $x \in U'$, (a) $\neg\xi(x,\alpha)!$, or (b) $\xi(x,tick)!$, or (c)
$\neg(\exists s' \in \Sigma^*)$ ($\xi(x_0,s')=x$, $\hat\xi(U_0, Ps') = U'$ and $\delta(q_0,s'.tick)!$).
First, Case (c) is impossible, because we already have $\xi(x_0,s) \in U'$, $\hat\xi(U_0, Ps) = U'$, and
$s.tick \in L({\bf G})$ (namely string $s$ falsifies the logical
statement of Case (c)). 
Next, Case (b) means directly that $s.tick \in L({\bf SUP})$.
Finally, Case (a) implies that $\alpha \notin Elig_{L_m({\bf SUP})}(s)$; note that this holds for
all $\beta \in \Sigma_{for}$. Hence $Elig_{L_m({\bf SUP})}(s) \cap \Sigma_{for} = \emptyset$.
Then by the fact that $L_m({\bf SUP})$ is controllable and $s.tick \in L({\bf                                                                                                                            G})$, $tick \in Elig_{L_m({\bf SUP})}(s)$, i.e. $s.tick \in L({\bf SUP})$.

(ii) $tick \in \Sigma_{o}$. In this case, for the state $i \in I$ of the generator ${\bf J}_\alpha$
corresponding to $y$ (i.e. $i = \zeta_\alpha(i_0, P(s))$), there holds $\zeta_\alpha(i,tick)!$.
By the definition of $\zeta_\alpha$ in (\ref{eq:sub3:construct}), there exists an uncertainty
set $U \in \mathcal{U}_i$ such that $\hat\xi(U,tick)!$. So $E_{tick}(U) = 1$. The rest of
the proof is identical to Case (i) above, and we conclude that $s.tick \in L({\bf SUP})$ as well.

The ($\supseteq$) direct of (\ref{eq:sub1:equv}), as well as
equation (\ref{eq:sub2:equv}) can be established similarly to
\cite{ZhangCW17}. \hfill $\square$

Finally, we provide the proof of Lemma~\ref{lem:preempt}.

{\it Proof of Lemma~\ref{lem:preempt}.} We must prove (\ref{eq:Tloc1}) and
(\ref{eq:Tloc2}).

First, for ($\Rightarrow$) of Eq. (\ref{eq:Tloc1}), let $P_\alpha(s).tick \in L({\bf LOC}_\alpha^P)$,
$s.tick \in L({\bf G})$ and $s\alpha \in L({\bf SUP})$; we must prove
that $s.tick \in L({\bf SUP})$. It is derived from $P_\alpha(s).tick \in L({\bf LOC}_\alpha^P)$
that $P_\alpha(s) \in L({\bf LOC}_\alpha^P)$, because $L({\bf LOC}_\alpha^P)$ is prefix-closed.
Let $y := \eta_\alpha(y_{0,\alpha},P_\alpha(s))!$; by $P_\alpha(s).tick \in L({\bf LOC}_\alpha^P)$,
$\eta_\alpha(y,tick)!$. The rest of the proof is identical
to the inductive case of proving ($\subseteq$) of (\ref{eq:sub1:equv}),
and we conclude that $s.tick \in L({\bf SUP})$.

Next, for ($\Leftarrow$) of Eq. (\ref{eq:Tloc1}), let $s.tick \in L({\bf SUP})$
and $s\alpha \in L({\bf SUP})$; $s\in L({\bf SUP})$ and $s.tick \in L({\bf G})$ are immediate, and it is left to show
that $P_\alpha(s).tick \in L({\bf LOC}_\alpha^P)$. By $s.tick \in L({\bf SUP})$
and (\ref{eq:sub1:equv}), we have for all $\sigma \in \Sigma_{for}$,
$s.tick \in P^{-1}_\sigma L({\bf LOC}_\sigma^P)$. Because $\alpha \in \Sigma_{for}$,
we have $s.tick \in P^{-1}_\alpha L({\bf LOC}_\alpha^P)$,
and thus $P_\alpha(s.tick) \in L({\bf LOC}_\alpha^P)$.
According to the definition of $\Sigma_\alpha$, $\{tick\} \subseteq \Sigma_\alpha$.
Hence, $P_\alpha(s).tick = P_\alpha(s.tick) \in L({\bf LOC}_\alpha^P)$.

Finally, to prove (\ref{eq:Tloc2}), let $y, y' \in Y_\alpha$
and assume that $y' = \eta_\alpha(y,\sigma)$ and $y \neq y'$; we prove that
$\sigma \in \Sigma_o$ by contradiction. Suppose that $\sigma \in \Sigma_{uo}$.
According to (\ref{eq:sub3:construct}), for all $i \in I$, $\zeta_\alpha(i,\sigma)$
is not defined. Further, according to the rules (iii) and (iv)
of constructing ${\bf LOC}_\alpha^P$, initially only the transitions
labeled by observable events are added to $\eta_\alpha$. Thus
for all $y\in Y$ and $\sigma \in \Sigma_{uo}$, $\eta_\alpha(y,\sigma)$ is not defined,
which contradicts the assumption that $y' = \eta_\alpha(y,\sigma)$.
Then, if $\alpha$ (resp. $tick$) is unobservable and $\psi_\alpha(y) = 1$ (resp. $\psi_{tick}(y) = 1$), then
$\alpha$-selfloops (resp. $tick$-selfloops) are added to $\eta_\alpha$. Namely, only
the selfloops $\eta_\alpha(y,\alpha) = y$ (resp. $\eta_\alpha(y,tick) = y$) are added to $\eta_\alpha$,
which contradicts the assumption that $y \neq y'$.
So we conclude that $\sigma \in \Sigma_o$.
\hfill $\square$


\subsection{Case Study: Timed Workcell} \label{subsec:casestudy1}

\begin{figure}[!t]
\centering
    \includegraphics[scale = 0.18]{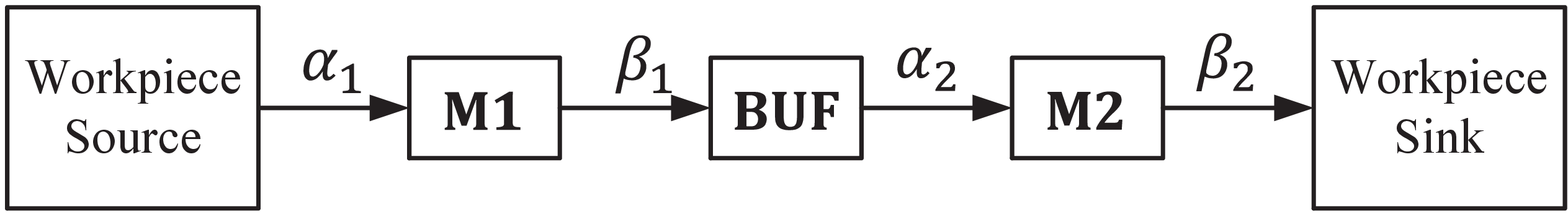}
\caption{Workcell: system configuration} \label{fig:WorkCell}
\end{figure}

\begin{figure}[!t]
\centering
    \includegraphics[scale = 0.20]{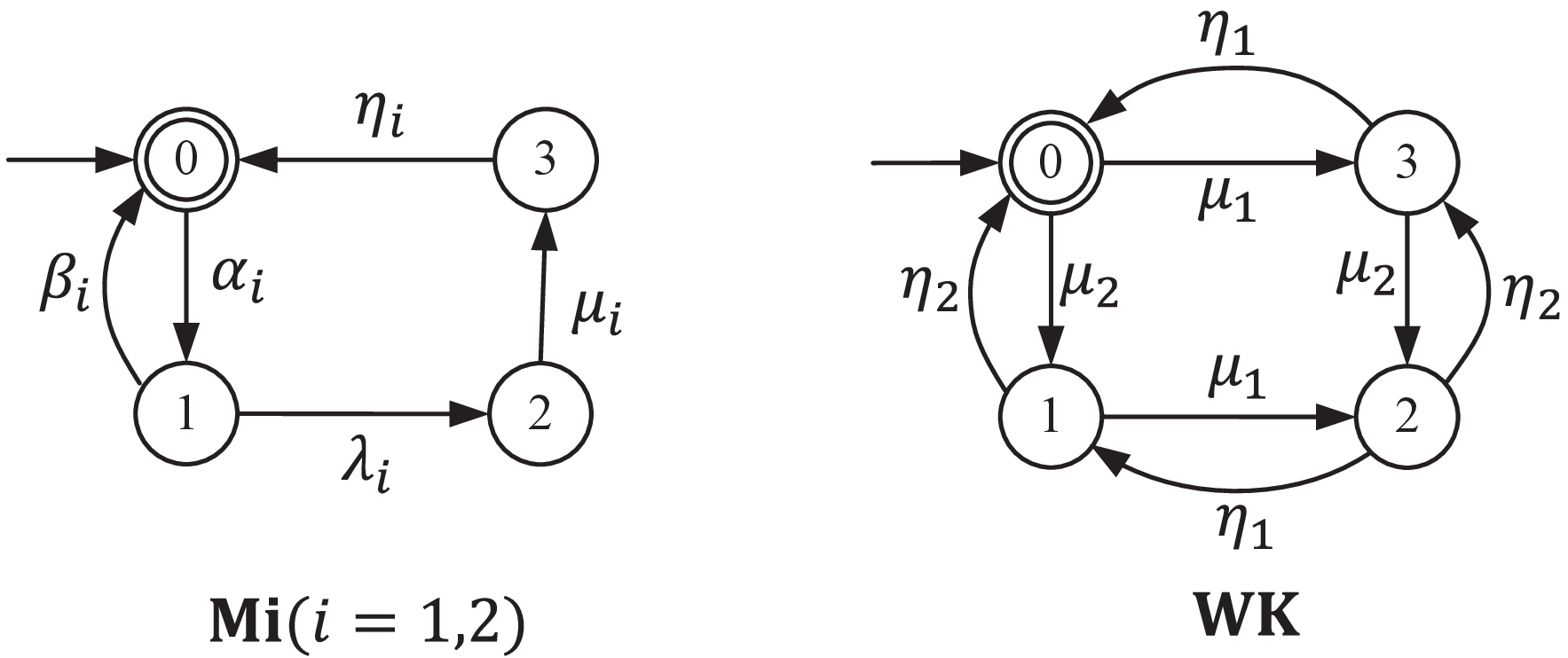}
\caption{ATG of plant components} \label{fig:machine}
\end{figure}

We illustrate the proposed partial-observation supervisor localization procedure by a timed workcell
example, adapted from \cite[Chapter 9]{Wonham16a}. As displayed in Fig.~\ref{fig:WorkCell},
the workcell consists of two machines $\bf M1$ and $\bf M2$, linked by a one-slot buffer $\bf BUF$;
additionally, a worker $\bf WK$ is responsible for repairing $\bf M1$ and $\bf M2$.
The ATG of the machines and the worker are displayed in Fig.~\ref{fig:machine}. The workcell
operates as follows. Initially the buffer is empty. With the event $\alpha_1$, $\bf M1$ takes
a workpiece from the infinite workpiece source. Subsequently $\bf M1$ either breaks down (event
$\lambda_1$), or successfully completes its work cycle, deposits the workpiece in the buffer
(event $\beta_1$). $\bf M2$ operates similarly, but takes its workpiece from the buffer (event
$\alpha_2$), and deposits it when finished in the infinite workpiece sink. If a machine $\bf Mi$, $i = 1$ or $2$
breaks down (event $\lambda_i$), then the worker $\bf WK$ will start to repair the machine (event $\mu_i$),
and finish the repair (event $\eta_i$) in due time.
Assign lower and upper time bounds to each event, with notation (event, lower bound, upper bound), as follows:
\begin{align*}
{\bf M1}&\text{'s timed events}: \\
&(\alpha_{1},0,\infty)~~(\beta_{1},1,2)~~(\lambda_{1},0,2)~~(\mu_{1},0,\infty)~~(\eta_{1},1,\infty)\\
{\bf M2}&\text{'s timed events}: \\
&(\alpha_{2},0,\infty)~~(\beta_{2},1,1)~~(\lambda_{2},0,1)~~(\mu_{2},0,\infty)~~(\eta_{2},2,\infty)\\
{\bf WK}&\text{'s timed events}: \\
&(\mu_{1},0,\infty)~~(\eta_{1},1,2)~~(\mu_{2},0,\infty)~~(\eta_{2},2,3)
\end{align*}

Then the TDES models of the two machines and the worker can be generated \cite{Wonham16a}; their joint behavior
is the composition of the three TDES, which is the plant $\bf PLANT$ to be controlled, i.e.
\[{\bf PLANT} = {\bf Comp}({\bf M1}, {\bf M2}, {\bf WK}).\]
Note that $\bf Mi$ ($i = 1, 2$) shares events $\mu_i$ and $\eta_i$ with $\bf WK$; so according to the
composition rule described in Section II, the lower and upper bounds of $\mu_i$ and $\eta_i$
are unified as:
$(\mu_{1},0,\infty)~~(\eta_{1},1,2)~~(\mu_{2},0,\infty)~~(\eta_{2},2,3)$.

\begin{figure}[!t]
\centering
    \includegraphics[scale = 1.0]{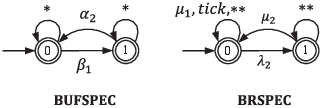}
\caption{Control specifications: $* = \{tick,\alpha_1,\lambda_1,\mu_1,\eta_1,\beta_2,\lambda_2,\mu_2,\eta_2\}$,
and $** = \{\alpha_1,\beta_1,\lambda_1,\eta_1,\alpha_2,\beta_2,\eta_2\}$} \label{fig:Specs}
\end{figure}

To impose behavioral constraints on the two machine's joint behavior, we take $\Sigma_{for} = \Sigma_{hib}
= \{\alpha_i,\mu_i|i = 1, 2\}$, and $\Sigma_{uc} = \{\beta_i, \lambda_i, \eta_i|i = 1, 2\}$. We impose
the following control specifications: (S1) $\bf BUF$ must not overflow or underflow; (S2) if $\bf M2$ goes
down, its repair must be started ``immediately'', and prior to starting repair of $\bf M1$ if $\bf M1$
is currently down. These two specifications are formalized as generators $\bf BUFSPEC$ and $\bf BRSPEC$ respectively,
as displayed in Fig.~\ref{fig:Specs}. So the overall specification imposed on the $\bf PLANT$
is represented by ${\bf SPEC} = {\bf BUFSPEC} || {\bf BRSPEC}$, where `$||$' denotes the synchronous
product of two generators \cite{Wonham16a}.

\begin{figure}[!t]
\centering
    \includegraphics[scale = 1.1]{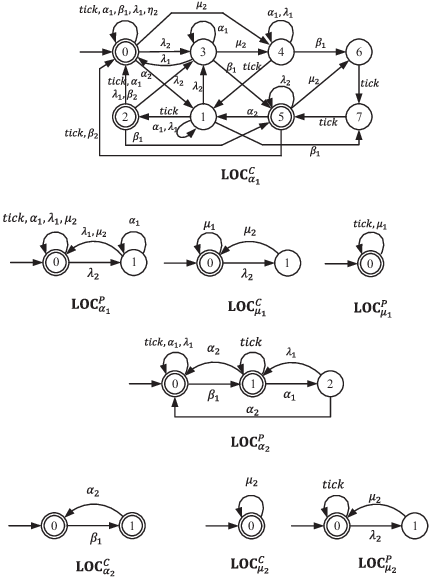}
\caption{Local preemptors and local controller under partial observation with
$\Sigma_{uo} = \{\mu_1,\eta_2\}$} \label{fig:relobs_locs}
\end{figure}

For partial observation we set $\Sigma_{uo}=\{\mu_1,\eta_2\}$,
namely the event of starting repair of $\bf M1$ and event of
finishing the repair of $\bf M2$ are unobservable. Note
that $\mu_1$ is both prohibitible and forcible, while $\eta_2$
is uncontrollable. We first compute as in (\ref{eq:monosup})
the controllable and observable behavior $\bf SUP$, which
has 77 states and 169 transitions. Then we apply the proposed partial-observation supervisor
localization procedure to construct partial-observation local preemptors and partial-observation local
controllers, respectively for each forcible event and each prohitibile event. The computation
is done by an algorithm adapted from \cite{ZhangCW17}, as discussed in Section~\ref{subsec:mainresult}.
The results are displayed in Fig.~\ref{fig:relobs_locs}; it is inspected from the TTG of
the local preemptors/controllers that none of the unobservable events (in $\Sigma_{uo}=
\{\mu_1, \eta_2\}$) causes state change. It is also verified that the collective
controlled behavior of these local preemptors and controllers is identical to the controllable
and observable behavior ${\bf SUP}$. In the following we
explain the control logics of the constructed local preemptors and
controllers.

Local controller ${\bf LOC}_{\alpha_1}^C$ guarantees no overflow of the buffer.
There are two cases that are safe for $\bf M1$ to take a workpiece from the source (i.e. executing event $\alpha_1$).
First, no workpiece has been deposited into the buffer (${\bf LOC}_{\alpha_1}^C$ at state 0),
or a deposited workpiece has been taken away by $\bf M2$ (${\bf LOC}_{\alpha_1}^C$ at states 1, 2, 3, 4).
Second, the buffer is full and $\bf M2$ is ready to take one workpiece from the buffer
(${\bf LOC}_{\alpha_1}^C$ returns to state 0). Since the lower bound of $\beta_1$ and
$\lambda_1$ is 1, $\bf M1$ will either complete
a cycle ($\alpha_1\beta_1$), or break down when one $tick$ passes after it takes a workpiece from the source, and before that $tick$
occurs, can $\bf M2$ effectively (via preempting $tick$ event) take a workpiece from the buffer. Namely, the buffer will be empty (because $\bf M2$ has taken away the workpiece) before $\bf M1$
deposits the workpiece into it. So at this time it is safe for $\bf M1$ to take a workpiece from the buffer. In other cases (states 5, 6, and 7 of ${\bf LOC}_{\alpha_1}^C$),
event $\alpha_1$ must be prohibited.
On the other hand, local preemptor ${\bf LOC}_{\alpha_1}^P$ describes that the occurrence of $\alpha_1$ may
preempt $tick$ event when $\bf M2$ breaks down. The reason is as follows.
$\bf M2$ may break down only after it has taken a workpiece from
the buffer. Thus at this time the buffer is empty, and
it is safe for $\bf M1$ to take a workpiece from the source. According to specification (S2), however, the repair of $\bf M2$ must be started immediately.
Hence, before $\bf WK$ starts to repair $\bf M2$, the occurrence of $\alpha_1$
must preempt the $tick$ event.
Note that this logic does not violate specification (S2), because $\mu_2$ is only allowed to preempt
event $tick$, but not any other events.

${\bf LOC}_{\mu_1}^C$ disables event $\mu_1$ when $\bf M2$ breaks down, as required by
specification (S2), i.e. the starting repair of $\bf M2$ is prior to that of $\bf M1$.
${\bf LOC}_{\mu_1}^P$ describes a preemption logic that the starting repair of $\bf M1$
need not preempt $tick$ event.

${\bf LOC}_{\alpha_2}^C$ guarantees no underflow of the buffer. Only after $\bf M1$
has deposited a workpiece into the buffer will $\bf M2$ takes the workpiece.
The logic of ${\bf LOC}_{\alpha_2}^P$ is to preempt $tick$ when the buffer is full
and $\bf M1$ has taken a workpiece from the source.

${\bf LOC}_{\mu_2}^C$ and ${\bf LOC}_{\mu_2}^P$ ensure no violation of specification (S2).
First, according to ${\bf LOC}_{\mu_2}^C$, $\mu_2$ is enabled all the time because the repair of $\bf M2$ has higher priority than that of $\bf M1$. Second, according to ${\bf LOC}_{\mu_2}^P$
the repair of $\bf M2$ must be started immediately
if it breaks down, which effectively preempts event $tick$.

\begin{figure}[!t]
\centering
    \includegraphics[scale = 0.8]{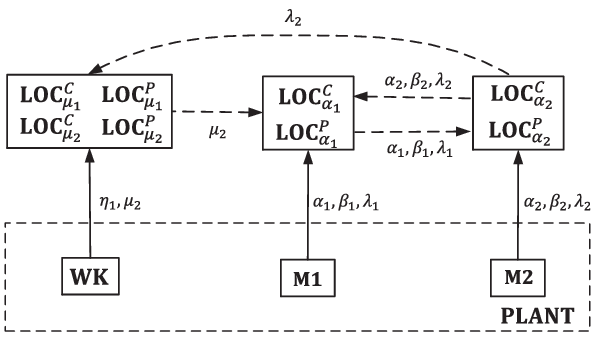}
\caption{Distributed control architecture under partial observation with
$\Sigma_{uo} = \{\mu_1, \eta_2\}$} \label{fig:DisComDiag}
\end{figure}

Finally, according to the allocation policy described in Section~\ref{sec:ProbFormu},
we build a distributed control architecture for the timed workcell, as displayed in
Fig.~\ref{fig:DisComDiag}.
Here $\hat\Sigma_{for,{\bf WK}} = \hat\Sigma_{hib,{\bf WK}} = \{\mu_1,\mu_2\}$,
$\hat\Sigma_{for,{\bf M1}} =
\hat\Sigma_{hib,{\bf M1}} = \{\alpha_1\}$, and $\hat\Sigma_{for,{\bf M2}} = \hat\Sigma_{hib,{\bf M2}} = \{\alpha_2\}$.
A local preemptor/controller either
directly observes an observable event generated by the plant component owning it, as denoted by solid lines
in Fig.~\ref{fig:DisComDiag}, or imports an observable event by communication from other
local preemptors/controllers, as denoted by the dashed lines. Those
events imported by communication may be subject to delay when using
physical channels; we shall address this problem in the next section.

\section{Supervisor Localization of TDES with Communication Delay} \label{sec:delay}


By the supervisor localization procedure presented in Section IV and the
allocation policy described in Section \ref{sec:ProbFormu}, we have built a distributed
control architecture for TDES $\bf G$ in (\ref{eq:multi_agent}). Each agent
${\bf G}_k$ ($k \in \mathcal{N}:=\{1,...,N\}$) \footnote{Since each agent corresponds to
exactly one plant component, without confusion we also use ${\bf G}_k$ to denote
the agent corresponding to the plant component ${\bf G}_k$.} owns a set of partial-observation local
preemptors  ${\bf LOC}_\alpha^P$ ($\alpha \in \hat\Sigma_{for,k}$), each with
a communication event set $\Sigma_{com,\alpha}$ (as in (\ref{eq:comevent_prmpt})),
and a set of partial-observation local controllers ${\bf LOC}_\beta^C$
($\beta\in \hat\Sigma_{hib,k}$), each with a communication event set $\Sigma_{com,\beta}$ (as in
(\ref{eq:comevent_disab})). So far it has been assumed that there is
no delay of event communication.

In this section we consider that the events in the communication sets are transmitted
through physical channels and thus subject to (generally non-zero) communication delays.
Specifically, consider that one of the communication events, say $\sigma$, is transmitted
from agent ${\bf G}_i$ to ${\bf G}_j$ through some communication media and with non-zero delay. Physically, the occurrence
of event $\sigma$ is  observable only by the sender ${\bf G}_i$,
but not by the receiver ${\bf G}_j$. Instead, through the communication media,
${\bf G}_j$ will receive the occurrence of $\sigma$ after some time (i.e. communication delay).
Denote the event of receiving $\sigma$ by a new event label $\sigma'$;
thus $\sigma'$ is observable by the receiver ${\bf G}_j$ (but not by the sender ${\bf G}_i$).
As a result of delay, ${\bf G}_i$ and ${\bf G}_j$ has distinct observable
event sets, and they must take their preemptive/control actions accordingly.


\subsection{Communication Channel Models} \label{subsec:chnmodel}


Let ${\bf G}_l$ ($l \in \mathcal{N}$) be an agent with event set $\Sigma_l$, and
denote by $\Sigma_{com,l}$ the subset of events to be communicated to ${\bf G}_l$
which is given by
\begin{align*}
\Sigma_{com,l} = ~\big(\mathop \bigcup\limits_{\alpha \in \hat\Sigma_{for,l}}(\Sigma_{com,\alpha}\setminus &\Sigma_l)\big) \notag\\
\cup~~ &\big(\mathop \bigcup\limits_{\beta \in \hat\Sigma_{hib,l}}(\Sigma_{com,\beta}\setminus \Sigma_l)\big).
\end{align*}
Let ${\bf G}_k$ ($k \in \mathcal{N}$) be another agent with $\Sigma_k$.
Then the subset of events communicated from agent ${\bf G}_k$ to ${\bf G}_l$ is
\begin{align} \label{eq:kcoml}
\Sigma_{k,com,l} = \Sigma_{k} \cap \Sigma_{com,l}.
\end{align}

In the following we focus on non-zero communication delays, and
represent by $\Sigma_{k,com,l}' \subseteq \Sigma_{k,com,l}$ the subset of events whose communication delays are greater than zero. Those events in $\Sigma_{k,com,l}\setminus \Sigma_{k,com,l}'$ are transmitted with no delay, and thus can be observed directly by the receiver; hence in this case we do not employ channel models for their transmissions.

\begin{figure}[!t]
\centering
    \includegraphics[scale = 1.0]{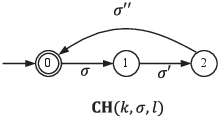}
\caption{ATG of TDES channel model ${\bf CH}(k,\sigma,l)$}
\label{fig:channel}
\end{figure}

Fix an event $\sigma \in \Sigma_{k,com,l}'$.
We propose a TDES channel model ${\bf CH}(k,\sigma,l)$, as displayed in
Fig.~\ref{fig:channel}.
In ${\bf CH}(k,\sigma,l)$, (i) event $\sigma$ denotes that $\sigma$ occurs in
${\bf G}_k$ and is sent to the communication channel; (ii) event $\sigma'$ denotes that
$\sigma$ is received by ${\bf G}_l$, and an acknowledgement message
is sent back to the channel; (iii) event $\sigma''$ denotes
that ${\bf G}_k$ receives the acknowledgement, which simultaneously resets the
channel to be idle (i.e. the channel is ready to send the next
occurrence of $\sigma$). The lower and upper bounds of events
$\sigma'$ and $\sigma''$ are determined by the practical requirements
on the communication delay bounds of $\sigma$. Note that the events $\sigma'$ and
$\sigma''$ are specific to ${\bf CH}(k,\sigma,l)$, which transmits event $\sigma$
from ${\bf G}_k$ to ${\bf G}_l$. In other words, if we adopt another channel ${\bf CH}(k,\sigma,l')$
to transmit event $\sigma$ from ${\bf G}_k$ to ${\bf G}_{l'}$, we will use other notation, e.g. $\hat\sigma'$ and $\hat\sigma''$,
to replace $\sigma'$ and $\sigma''$ respectively.
Here for simplicity we adopt $\sigma'$ and $\sigma''$ in ${\bf CH}(k,\sigma,l)$ as a generic case.

First, to meet a hard deadline of an operation or to ensure system's
timely performance in practice, it may often be the case that
the communication delay of event $\sigma$ is bounded by $d \in \mathbb{N}-\{0\}$ $tick$s.
In this case, the lower time bounds of $\sigma'$ and $\sigma''$ are both
set to be $0$ and the upper bounds to be $d$, which means
that the time consumed for transmitting the occurrence of $\sigma$ from ${\bf G}_k$ to ${\bf G}_l$
{\it and} that for acknowledging the receival of $\sigma$ from ${\bf G}_l$ to ${\bf G}_k$ should be both
no more than $d$ $tick$s.

Second, in case there happens to be no specific deadline requirement on transmission
of event $\sigma \in \Sigma_{k,com,l}'$, or simply no {\it a priori} knowledge is
available of a delay bound on $\sigma$, it may be reasonable to consider {\it unbounded
delay} of $\sigma$-communication. This means that the transmission of $\sigma$ may take
{\it indefinite} time to complete, although it will complete eventually. So, in this
case, $\sigma'$ and $\sigma''$ both have lower bound $0$ and upper bound $\infty$
(i.e. they may occur at any time after they become eligible to).

\begin{figure}[!t]
\centering
    \includegraphics[scale = 1.0]{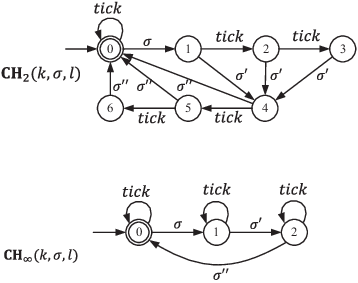}
\caption{TTG of bounded channel model ${\bf CH}_d(k,\sigma,l)$ with delay $d = 2$ and unbounded channel
model ${\bf CH}_\infty(k,\sigma,l)$; the TTG are obtained from the ATG displayed in Fig.~\ref{fig:channel}
with $(l_\sigma,u_\sigma) = (0,2)$ and $(l_\sigma,u_\sigma) = (0,\infty)$ respectively
(by applying the constructing rules in \cite[Chapter 9]{Wonham16a}). In the models, first, the
occurrence of event $\sigma$ means that $\sigma$ occurs in
${\bf G}_k$ and is sent to the communication channel; after some time delay (less than $d$),
$\sigma'$ will occur, which represents that the occurrence of $\sigma$ is received by ${\bf G}_l$, and an acknowledgement message
is sent back to the channel; finally after another time delay the occurrence of event $\sigma''$ denotes
that ${\bf G}_k$ receives the acknowledgement, which simultaneously resets the
channel to be idle. }
\label{fig:chnmodel}
\end{figure}

To distinguish the channel models ${\bf CH}(k,\sigma,l)$ of the above two cases, in notation
we use ${\bf CH}_d(k,\sigma,l)$ to represent the channel with delay bound $d$,
and ${\bf CH}_\infty(k,\sigma,l)$ the channel with unbounded delay.
An example of bounded channel model of $\sigma$ with delay bound $d = 2$ and unbounded
channel model is given in Fig.~\ref{fig:chnmodel}.




In the channel models above, we make the following choices. (i) Both events
$\sigma'$ and $\sigma''$ are uncontrollable, because it is not reasonable
(if not impossible) to disable the receipt of a communication or an acknowledgement;
(ii) events $\sigma$, $\sigma''$ are observable to the sender ${\bf G}_k$ but unobservable
to the receiver ${\bf G}_l$, while $\sigma'$ is observable to ${\bf G}_l$ but unobservable to
${\bf G}_k$. This means that the agents ${\bf G}_1$,...,${\bf G}_N$ generally have different
subsets of observable events; this is a new feature of the current formulation with
communication delay. Intuitively, to obtain local preemptors/controllers in this formulation,
we need to iteratively apply the supervisor localization under the different partial observations; this
can be realized by combining timed relative coobservability \cite{CaiZW16} with supervisor localization, as will be described as follows.

\begin{remark}
The communication channel models proposed above differ from those in \cite{ZhangEt16b} in
the following two respects. First, the models in this paper are richer with adding
an event label $\sigma'$ to represent that the receiver has received the occurrence
of event $\sigma$ in the sender and sent an acknowledgement back to the channel.
By this operation, the communication delay in transmitting the occurrence of $\sigma$ and
that in transmitting the acknowledgement information are modeled separately, while in \cite{ZhangEt16b} the communication delays are accumulated
as a single value. Hence the models in this paper are more practical. Second, the channel models in this paper are considered as plant components
and they together with the original components form the new plant to be controlled; namely,
the delays will be considered as part of plant dynamics in the supervisor synthesis procedure. While in \cite{ZhangEt16b} the delays are not considered
in the supervisor synthesis procedure, and thus it is not guaranteed that the synthesized supervisors can tolerate the given delays.
\end{remark}


\subsection{Partial-Observation Supervisor Localization with Communication Delay} \label{subsec:delaymainresult}


Recall from Section III that, in the delay-free case, we had
plant ${\bf G} = {\bf Comp} ({\bf G}_1, ..., {\bf G}_N)$ over $\Sigma$,
specification $E \subseteq \Sigma^*$, prohibitible event set $\Sigma_{hib}$,
and forcible event set $\Sigma_{for}$. 

Now for $k,l \in \mathcal{N}$ let $\Sigma_{k,com,l}'$ be partitioned as
$\Sigma_{k,com,l}' = \Sigma_{k,com,l}^{bd} \dot{\cup} \Sigma_{k,com,l}^{ud}$, where
$\Sigma_{k,com,l}^{bd}$
is the subset of communication events with bounded
delay and $\Sigma_{k,com,l}^{ud}$ the subset of those with unbounded delay.
First, the new plant $\tilde{\bf G}$ including both the plant components of $\bf G$ and the channels is
\begin{align} \label{eq:newplant}
\tilde{\bf G} = {\bf Comp}(&{\bf G},\{{\bf CH}(k,\sigma,l)|\sigma
\in \Sigma_{k,com,l}^{bd}, k, l \in \mathcal{N}\}, \notag \\
\{&{\bf CH}(k,\sigma,l)|\sigma \in \Sigma_{k,com,l}^{ud}, k, l \in \mathcal{N}\}),
\end{align}
where ${\bf CH}(k,\sigma,l)$ is the ATG displayed in Fig. \ref{fig:channel}.
The event set $\tilde{\Sigma}$ of $\tilde{\bf G}$ is $\tilde{\Sigma}
= \Sigma ~\cup ~\{\sigma',\sigma''|\sigma \in \Sigma_{k,com,l}', k,l \in \mathcal{N}\}$.
Since none of the added events $\sigma'$ and $\sigma''$ is
forcible, or prohibitible, the new subset of forcible events and prohibitible
events are unchanged, i.e. $\tilde\Sigma_{for} = \Sigma_{for}$ and
$\tilde{\Sigma}_{hib} = \Sigma_{hib}$. So $\hat\Sigma_{for,k}$ and $\hat\Sigma_{hib,k}$
(as defined in (\ref{eq:disforce})) are also unchanged. Following the allocation policy for building distributed
control architecture, we choose $\hat\Sigma_{for,k}$ (resp. $\hat\Sigma_{hib,k}$)
to be the subset of forcible (resp. prohibitible) events for component ${\bf G}_k$ in
the new plant, i.e.,
\begin{align}
\tilde{\Sigma}_{for,k} &:= \hat\Sigma_{for,k} \label{eq:new_forci_event_k}\\
\tilde{\Sigma}_{hib,k} &:= \hat\Sigma_{hib,k} \label{eq:new_prohib_event_k}
\end{align}
Since $\hat\Sigma_{for,k}$ and $\hat\Sigma_{hib,k}$, ($k \in \mathcal{N}$)
are pairwise disjoint, so are  $\tilde\Sigma_{for,k}$ and $\tilde\Sigma_{hib,k}$.
Therefore, $\tilde\Sigma_{hib} = \dot{\bigcup}_{k \in \mathcal{N}} \tilde\Sigma_{hib,k}$
and $\tilde\Sigma_{for} = \dot{\bigcup}_{k \in \mathcal{N}} \tilde\Sigma_{for,k}$.

The specification imposed on $\bf G$ is not changed, but should be extended
to the new event set $\tilde\Sigma$, i.e. the specification
$\tilde{E} = \tilde{P}^{-1} E$,
where $\tilde{P}:\tilde\Sigma^*\rightarrow\Sigma^*$ is the natural projection.

As we have mentioned, a consequence of
introducing the communication channels is that the agents ${\bf G}_k$ ($k \in \mathcal{N}$)
have distinct observable event sets. Hence the local preemptors/controllers to be allocated to different
agents will be required to have different observable event sets. To address this, rather than synthesizing
a monolithic supervisor for a single observable event subset $\Sigma_o$, we propose to synthesize $N$ decentralized
supervisors one for each observable event set $\tilde\Sigma_{o,k}$ ($k \in \mathcal{N}$) given by
\begin{align*}  
\tilde\Sigma_{o, k} := ({\Sigma_{o} \setminus \Sigma_{com, k}'} ) &\cup \{\sigma,\sigma''|\sigma \in \Sigma_{k,com,l}', l \in \mathcal{N}, l \neq k\} \\
&\cup \{\sigma'|\sigma \in \Sigma_{l,com,k}', l \in \mathcal{N}, l \neq k\}.
\end{align*}

 For the synthesis of decentralized supervisors,
it is proved in \cite{RudWon:1992,ParkChoi09} that a set of decentralized supervisors exists
which synthesizes a language $K\subseteq L_m({\bf G})$ if and only if $K$
is coobservable, controllable and $L_m({\bf G})$-closed. Like observability, coobservability is not closed under
set union; consequently when $K$ is not coobservable, there generally does not exist the supremal coobservable
(and controllable, $L_m({\bf G})$-closed) sublanguage of $K$, and there is no
existing algorithm that computes a coobservable sublanguage of $K$. In this paper, we employ the concept
of (timed) relative coobservability \cite{CaiZW16}, which is stronger
than coobservability (thus only a sufficient condition for existence of
decentralized supervisors), but the supremal (timed) relatively coobservable sublanguage always exists.
Let $\tilde P_k: \tilde\Sigma^* \rightarrow \tilde\Sigma^*_{o,k}$ and $C \subseteq L_m(\tilde{\bf G})$
be an ambient language. A sublanguage $K \subseteq C$ is {\it timed relatively coobservable} (with respect to $C$, $\tilde{\bf G}$
and $\tilde P_k$, $k \in \mathcal{N}$), or simply timed $C$-coobservable, if
for every $k \in \mathcal{N}$ and
every pair of strings $s,s' \in \Sigma^*$ with $\tilde P_k (s) = \tilde P_k (s')$
there holds
\begin{align*}
&(\forall \sigma \in{\Sigma_{k,act}\cup\{tick\}}) \\
 &~~~~~~~~~~~~~s\sigma \in \overline{K}, s' \in \overline{C},
s'\sigma \in L(\tilde{\bf G}) \Rightarrow s'\sigma \in \overline{K}.
\end{align*}
Namely, relative coobservability of $K$ requires that $K$ be relatively observable with
respect to each $\tilde P_{o,k}$ and $\Sigma_k$, $k \in \mathcal{N}$.
It is proved in
\cite{CaiZW16} that there always exists a unique supremal
relatively coobservable sublanguage of a given language, which may be effectively computed  by an algorithm in \cite{CaiZW16}. Since relative coobservability is stronger
than coobservability, the supremal relatively coobservable (and controllable, $L_m({\bf G})$-closed) sublanguage is guaranteed
to be coobservable (and controllable, $L_m({\bf G})$-closed), and
thereby ensures the existence of decentralized supervisors \cite{RudWon:1992,ParkChoi09}. 

For the new plant $\tilde{\bf{G}}$ and specification language $\tilde{E}$,
write $\mathcal{CCO}(\tilde{E} \cap L_m(\tilde{\bf{G}}))$ for the
family of relatively coobservable (and controllable, $L_m(\tilde{\bf
G})$-closed) sublanguages of $\tilde{E} \cap L_m(\tilde{\bf{G}})$.
Then $\mathcal{CCO}(\tilde{E} \cap L_m(\tilde{\bf{G}}))$ is nonempty
(the empty language $\emptyset$ belongs) and has a unique
supremal element
\[\sup\mathcal{CCO}(\tilde{E} \cap L_m(\tilde{\bf{G}})) = \bigcup\{K|K\in\mathcal{CCO}(\tilde{E} \cap L_m(\tilde{\bf{G}}))\}.\]
Let the generator ${\bf NSUP}$ be such that
\begin{equation} \label{eq:new_monosup}
L_m({\bf NSUP}) := \sup \mathcal {CCO}(\tilde{E} \cap L_m(\tilde{\bf{G}})).
\end{equation}
We call ${\bf NSUP}$ the {\it controllable and coobservable
behavior}, and assume that $L_m({\bf NSUP}) \neq \emptyset$.\footnote{The introduced
bounded/unbounded communication delays may cause $L_m({\bf NSUP}) = \emptyset$,
which means that the delay requirements are too strong to be satisfied. In that case,
we shall weaken the delay requirements by either decreasing delay bounds of bounded-delay channels
(when the delay bound of an event $\sigma$ needs to be decreased to 0, we do not employ a channel model
for $\sigma$, and consequently events $\sigma'$ and $\sigma''$ defined in the channel model are also removed from the alphabet)
or reducing the number of unbounded-delay channels, until we
obtain a nonempty $L_m({\bf NSUP})$. }

Next, for each observable event set $\tilde\Sigma_{o,k}$ ($k \in \mathcal{N}$),
we construct as in (\ref{eq:posup}) a {\it partial-observation decentralized supervisor}
${\bf NSUPO}_k$ defined over $\tilde\Sigma_{o,k}$. It is well-known \cite{RudWon:1992,LinWon95}
that such constructed decentralized supervisors ${\bf NSUPO}_k$
collectively achieve the same controlled behavior as $\bf NSUP$
does.
The control actions of the supervisor ${\bf NSUPO}_k$
include (i) preempting event $tick$ via forcible events in $\tilde\Sigma_{for,k}$ (as in (\ref{eq:new_forci_event_k}))
and (ii) disabling prohibitible events in $\tilde\Sigma_{hib,k}$ (as in (\ref{eq:new_prohib_event_k})).

Finally, we apply the localization procedure developed in Section IV to decompose,
one at a time, each decentralized supervisor ${\bf NSUPO}_k$, $k \in \mathcal{N}$.
The result is a set of partial-observation local preemptors ${\bf NLOC}_{\alpha}^P = (Y_{\alpha},
\Sigma_{\alpha},\eta_{\alpha}, y_{0,\alpha},Y_{m,\alpha})$, one for each forcible
event $\alpha \in \tilde\Sigma_{for}$, as well as a set of partial-observation local controllers ${\bf NLOC}_{\beta}^C = (Y_{\beta},
\Sigma_{\beta},\eta_{\beta}, y_{0,\beta},Y_{m,\beta})$, one for each $\beta \in \tilde\Sigma_{hib}$.
Owing to $\tilde\Sigma_{for} = \dot{\bigcup}_{k \in \mathcal{N}} \tilde\Sigma_{for,k}$
(resp. $\tilde\Sigma_{hib} = \dot{\bigcup}_{k \in \mathcal{N}} \tilde\Sigma_{hib,k}$),
one local preemptor ${\bf NLOC}_{\alpha}^P$ (resp. one local controller ${\bf NLOC}_{\beta}^C$)
will be owned by precisely one agent.

The following is the main result of this section, which asserts that the collective controlled
behavior of the resulting partial-observation local preemptors and local controllers, communicated
through the introduced channels with bounded/unbounded delays, is identical to that of {\bf NSUP}.

\begin{theorem} \label{thm:coobs_equ}
The set of partial-observation local preemptors $\{{\bf NLOC}_{\alpha}^P|\alpha \in
\tilde\Sigma_{for} \}$ and the set of partial-observation local
controllers $\{{\bf NLOC}_{\beta}^C|\beta \in \tilde\Sigma_{hib}\}$ derived above
are equivalent to the controllable and coobservable behavior ${\bf NSUP}$ in (\ref{eq:new_monosup}) with respect to
the plant $\tilde{\bf G}$, i.e.
\begin{align}
   L(\tilde{\bf G}) \cap L(\bf NLOC)  &= L(\bf NSUP) \label{eq:sub1:coobsequv}\\
   L_m(\tilde{\bf G}) \cap L_m(\bf NLOC)  &= L_m(\bf NSUP) \label{eq:sub2:coobsequv}
\end{align}
with
\begin{align}
L(\bf NLOC)~:=~ &\Big(\mathop \bigcap\limits_{\alpha \in \tilde\Sigma_{for}}P_{\alpha}'^{-1}L({\bf NLOC}^P_{\alpha}) \Big)\notag\\
           \cap~ &\Big(\mathop \bigcap\limits_{\beta \in \tilde\Sigma_{hib}}P_{\beta}'^{-1}L({\bf NLOC}^C_{\beta}) \Big) \label{eq:sub1:newloc}\\
L_m(\bf NLOC)~:=~ &\Big(\mathop \bigcap\limits_{\alpha \in \tilde\Sigma_{for}}P_{\alpha}'^{-1}L_m({\bf NLOC}^P_{\alpha})\Big)\notag\\
            \cap~ &\Big(\mathop \bigcap\limits_{\beta \in \tilde\Sigma_{hib}}P_{\beta}'^{-1}L_m({\bf NLOC}^C_{\beta}) \Big) \label{eq:sub2:newloc}
\end{align}
where $P_{\alpha}': \tilde\Sigma^* \rightarrow \Sigma_{\alpha}^*$
and $P_{\beta}': \tilde\Sigma^* \rightarrow \Sigma_{\beta}^*$.
\end{theorem}

The proof of Theorem~\ref{thm:coobs_equ}, presented below, is similar to that of Theorem
\ref{thm:equ}, which relies on the facts that (i) for each forcible
event, there is a corresponding partial-observation local preemptor that
preempts event $tick$ consistently with {\bf NSUP}, and (ii)
for each prohibitible event, there is a corresponding partial-observation
local controller that disables/enables it consistently with
${\bf NSUP}$. 

By the above localization approach, each agent ${\bf G}_k$ ($k \in \mathcal{N}$) acquires
a set of partial-observation local preemptors $\{{\bf NLOC}_{\alpha}^P | \alpha \in \tilde\Sigma_{for,k} \}$
and a set of partial-observation local controllers $\{{\bf NLOC}_{\beta}^C | \beta \in \tilde\Sigma_{hib,k} \}$.
Thus we obtain a distributed control architecture for multi-component TDES under partial observation and communication delay.

\vspace{2em}
\noindent {\it Proof of Theorem~\ref{thm:coobs_equ}}: The equality of (\ref{eq:sub2:coobsequv}) and the ($\supseteq$)
direction of (\ref{eq:sub1:coobsequv}) may be verified analogously
as in the proof of Theorem 1. Here we prove ($\subseteq$) of (\ref{eq:sub1:coobsequv})
by induction, i.e. $L(\tilde{\bf G}) \cap L({\bf NLOC}) \subseteq L({\bf NSUP})$.

For the {\bf base step}, note that none of $L(\tilde{\bf G})$, $L({\bf NLOC})$ and
$L({\bf NSUP})$ is empty; and thus the empty string $\epsilon$ belongs to all
of them. For the {\bf inductive step}, suppose that $s \in L(\tilde{\bf G}) \cap
L({\bf NLOC})$, $s \in L({\bf NSUP})$ and $s\sigma \in L(\tilde{\bf G}) \cap
L({\bf NLOC})$ for arbitrary event $\sigma \in \Sigma$; we must show that $s\sigma
\in L({\bf NSUP})$. Since $\tilde\Sigma = \tilde\Sigma_{uc}\dot\cup \tilde\Sigma_{hib}
\dot\cup\{tick\}$, we consider the following three cases.

(i) $\sigma \in \tilde\Sigma_{uc}$. Since $L({\bf NSUP})$ is controllable, and
$s\sigma \in L(\tilde{\bf G})$ (i.e. $\sigma \in Elig_{\tilde{\bf G}}(s)$), we have
$\sigma \in Elig_{L_m({\bf NSUP})}(s)$. That is, $s\sigma \in
\overline{L_m({\bf NSUP})} = L({\bf NSUP})$.

(ii) $\sigma = tick$. By the hypothesis that $s, s.tick \in L({\bf NLOC})$,
for every forcible event $\alpha \in \tilde\Sigma_{for,k}$, $k \in \mathcal{N}$,
$s, s.tick \in P_{\alpha}'^{-1}L({\bf NLOC}_{\alpha}^P)$, i.e. $P_{\alpha}'(s),
P_{\alpha}'(s).tick \in L({\bf NLOC}_{\alpha}^P)$. Let $y = \eta_\alpha(y_{0,\alpha},
P_{\alpha}'(s))$; then $\eta_\alpha(y,tick)!$. The rest of the proof is similar to
case (ii) of proving Theorem~\ref{thm:equ}, with ${\bf LOC}_{\alpha}^P$ and $P_\alpha$
replaced by ${\bf NLOC}_{\alpha}^P$ and $P_{\alpha}'$ respectively.

(iii) $\sigma \in \tilde\Sigma_{hib}$. There must exist a partial-observation local
controller ${\bf NLOC}_{\sigma}^C$ for $\sigma$. It follows from $s\sigma \in
L({\bf NLOC})$ that $s\sigma \in P_{\sigma}'^{-1}L({\bf NLOC}_{\sigma}^C)$ and
$s \in P_{\sigma}'^{-1}L({\bf NLOC}_{\sigma}^C)$. So $P_{\sigma}'(s\sigma)
\in L({\bf NLOC}_{\sigma}^C)$ and $P_{\sigma}'(s) \in L({\bf NLOC}_{\sigma}^C)$,
namely, $\eta_{\sigma}(y_0,P_{\sigma}'(s\sigma))!$ and $\eta_{\sigma}(y_0,
P_{\sigma}'(s))!$. Let $y := \eta_{\sigma}(y_0,P_{\sigma}'(s))$; then
$\eta_{\sigma}(y,\sigma)!$ (because $\sigma \in \Sigma_{\sigma}$).
The rest of the proof is similar to that
in \cite{ZhangCW17} for untimed DES. \hfill $\square$



\subsection{Case Study: Timed Workcell with Communication Delay}

We continue the timed workcell example in Section~\ref{subsec:casestudy1}
to illustrate the proposed partial-observation localization procedure with communication
delay. We assume that in the communication diagram Fig.~\ref{fig:DisComDiag}, the transmissions of the events $\beta_1$, $\lambda_1$ are subject to non-zero delay (at least one of these two events
must occur after $\bf M1$ has obtained a workpiece from the source).
For the communication delays, consider that
(i) event $\beta_1$ is transmitted from $\bf M1$ to $\bf M2$ with delay bound $d = 1$ ($tick$),
and (ii) event $\lambda_1$ is transmitted from $\bf M1$ to $\bf M2$ with unbounded delay bound.
The rest of the communication events are assumed (for simplicity) to be transmitted with no delay.
Continuing Section~\ref{subsec:casestudy1}, the events
$\mu_1$ and $\eta_2$ are unobservable.

\begin{figure}[!t]
\centering
    \includegraphics[scale = 0.9]{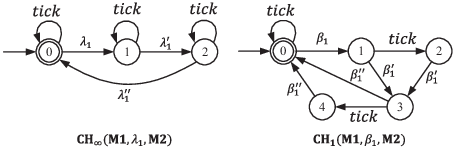}
\caption{TTG of TDES channel models ${\bf CH}_\infty({\bf M1},\lambda_1,{\bf M2})$,
and ${\bf CH}_1({\bf M1},\beta_1,{\bf M2})$. } \label{fig:CHNs}
\end{figure}

First, for event communications, we create TDES channel models
${\bf CH}({\bf M1},\beta_1,{\bf M2})$, and ${\bf CH}({\bf M1},\lambda_1,{\bf M2})$
to transmit events ${\beta_1}$ and ${\lambda_1}$, respectively. The lower and upper bounds of the newly added
events are listed in Table~\ref{tab:time_bound},
and the TTG of the channel models are displayed in Fig.~\ref{fig:CHNs}.

\begin{table}
\footnotesize
\caption{Time bounds of newly added events} \label{tab:time_bound}
\begin{center}
\scalebox{0.85}{
\begin{tabular}{|c|c||c|c|}
\hline
\multirow{2}{*}{event label} & (lower, upper)& \multirow{2}{*}{event label} & (lower, upper)\\
& bounds & & bounds \\
\hline
$\beta_1'$ & (0,1) & $\beta_1''$ & (0,1)  \\
\hline
$\lambda_1'$ & (0,$\infty$) & $\lambda_1''$ & (0,$\infty$)\\
\hline
\end{tabular}
}
\end{center}
\end{table}

Then, the new plant to be controlled is
\begin{align*}
{\bf NPLANT} = &{\bf Comp}({\bf M1}, {\bf M2},{\bf WK},  \\
                                       &~~{\bf CH}({\bf M1},\beta_1,{\bf M2}), {\bf CH}({\bf M1},\lambda_1,{\bf M2}))
\end{align*}
and the new specification is represented by $\bf NSPEC$, modified from $\bf SPEC$
(representing $E$) by adding selfloops of $\beta_1'$, $\beta_1''$, $\lambda_1'$ and $\lambda_1''$
 to each state of $\bf SPEC$ (as defined in Section~\ref{subsec:casestudy1}).
The subsets of observable events, forcible events and prohibitible events
are listed in Table~\ref{tab:new_eventset}. With these event sets, we compute the
controllable and coobservable controlled behavior ${\bf NSUP}$ as in (\ref{eq:new_monosup}),
which has 45 states and 78 transitions.

\begin{table}
\footnotesize
\caption{Subsets of observable, forcible, prohibitible events of
each component} \label{tab:new_eventset}
\begin{center}
\scalebox{0.85}{
\begin{tabular}{|c|c|c|c|}
\hline
\multirow{2}{*}{components} & \multirow{2}{*}{observable events} & forcible& prohibitible\\
                            &                                    & events  & events \\
\hline
\multirow{2}{*}{$\bf WK$} & $tick,\alpha_1,\beta_1,\lambda_1,\eta_1,$ & \multirow{2}{*}{$\mu_1,\mu_2$} & \multirow{2}{*}{$\mu_1,\mu_2$}  \\
&$\alpha_2,\beta_2,\lambda_2,\mu_2$&&\\
\hline
\multirow{2}{*}{$\bf M1$} & $tick,\alpha_1,\beta_1,\beta_1'', \lambda_1, \lambda_1'',\eta_1,$ & \multirow{2}{*}{$\alpha_1$} & \multirow{2}{*}{$\alpha_1$}  \\
&$\alpha_2,\beta_2,\lambda_2,\mu_2$&&\\
\hline
\multirow{2}{*}{$\bf M2$} & $tick,\alpha_1,\beta_1',\lambda_1', \eta_1,$ & \multirow{2}{*}{$\alpha_2$} & \multirow{2}{*}{$\alpha_2 $ }\\
&$\alpha_2,\beta_2,\lambda_2,\mu_2$&&\\
\hline
\end{tabular}
}
\end{center}
\end{table}

\begin{figure}[!t]
\centering
    \includegraphics[scale = 1.0]{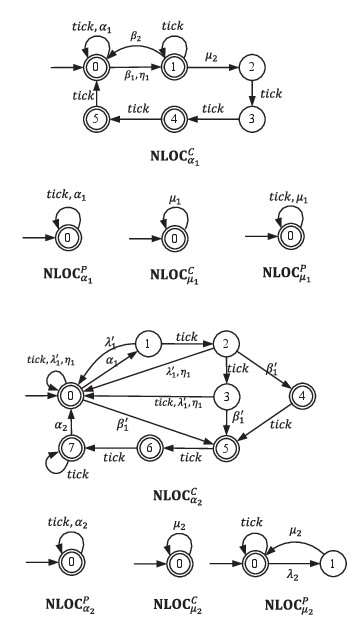}
\caption{Local preemptors and local controllers under partial observation and communication delays} \label{fig:relcoobs_locs}
\end{figure}

Next, we apply the proposed partial-observation supervisor localization procedure
presented in Section \ref{sec:LocProc} to construct a set of partial-observation local preemptors,
one for each forcible event in $\tilde\Sigma_{for}$ and a set of partial-observation local
controllers, one for each prohibitible event in $\tilde\Sigma_{hib}$. The results are displayed
in Fig~\ref{fig:relcoobs_locs}; it is inspected from the TTG of
the local preemptors/controllers that for the communication events transmitted
by the channels, only the events representing the receiving of an event occurrence (e.g.
$\beta_1'$ and $\lambda_1'$) cause state changes in the local controllers/preemptors
corresponding to the receivers. It is
verified that the collective controlled behavior of these local preemptors and controllers
is equivalent to ${\bf NSUP}$. The control logics of the partial-observation local
preemptors and controllers are similar to those described in Section~\ref{subsec:casestudy1}, but
affected by the communication delays. For illustration,
we consider the following two instances.

(i) Communication delays of $\beta_1$ and $\lambda_1$ affect the control logic of
${\bf NLOC}_{\alpha_2}^C$ and the preemptive logic of ${\bf NLOC}_{\alpha_2}^P$.
According to the control logic of ${\bf LOC}_{\alpha_2}^C$
described in Fig.~\ref{fig:relobs_locs}, $\bf M2$ will take a workpiece from the
buffer if it observes (event $\beta_1'$) that $\bf M1$ has deposited a workpiece into the buffer (event $\beta _1$).
However, now ${\bf NLOC}_{\alpha_2}^C$ cannot observe $\beta_1$ directly, and may know
(through the communication channels) the occurrence of event $\lambda_1$ before
that of $\beta_1$. Namely, it cannot judge which event of $\beta_1$ and $\lambda_1$ has occurred if
it does not receive their communicated events $\beta_1'$
and $\lambda_1'$, so the control logic of ${\bf NLOC}_{\alpha_2}^C$ becomes more
complicated: it will enable/disable event $\alpha_2$
according to the order of receiving of $\beta_1$ and $\lambda_1$.
Due to the change of ${\bf NLOC}_{\alpha_2}^C$, now the occurrence of
$\alpha_2$ will not preempt the event $tick$, as described by ${\bf NLOC}_{\alpha_2}^P$.

(ii) The communication delays of $\beta_1$ and $\lambda_1$ also affect the control logic of ${\bf NLOC}_{\alpha_1}^C$
and the preemptive logic of ${\bf NLOC}_{\alpha_1}^P$. As described
in (i), the occurrence of $\alpha_2$ cannot preempt event $tick$; thus
${\bf NLOC}_{\alpha_1}^C$ will enable event $\alpha_1$ only when the buffer is
empty (the plant is at the initial state or the workpiece in the buffer has
been taken away). This change also causes that the occurrence of $\alpha_1$
need not preempt event $tick$.

\begin{figure}[!t]
\centering
    \includegraphics[scale = 0.92]{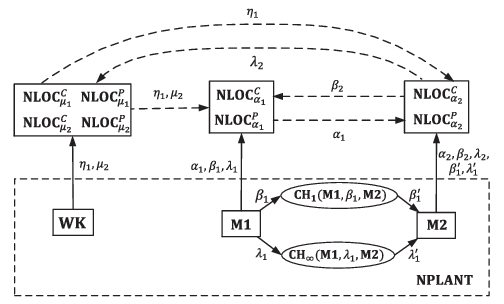}
\caption{Distributed control architecture with communication delay. 
} \label{fig:DisComDiag_Delay}
\end{figure}

Finally, by the same allocation policy applied to the delay-free case in Section~\ref{subsec:casestudy1},
we allocate the obtained local controllers and preemptors to the plant components $\bf M1$,
$\bf M2$, and $\bf WK$, thereby building a distributed control architecture under partial observation
and communication delay for the timed workcell, as displayed in Fig.~\ref{fig:DisComDiag_Delay}. A local
preemptor/controller may observe directly an event from the agent owning it, and
import an event from other agent through communication channels subject
to delay. Note that we selected for simplicity only two communication events ($\beta_1$ and $\lambda_1$)
to be transmitted through channels. By the same procedure
described above, however, one may easily add more communication
events transmitted through channels (i.e. by creating new
channel models and then applying the localization procedure
with communication delay again).

\section{Conclusions} \label{sec:concl}

In this paper, we have first developed a partial-observation supervisor
localization procedure to solve the distributed control problem of
multi-component TDES. A synthesized monolithic supervisor is decomposed
into a set of partial-observation local controllers and a set of
partial-observation local preemptors, whose state changes are caused
only by observable events. We have proved that the resulting local
controllers/preemptors collectively achieve the same controlled behavior
as the monolithic supervisor does.

Moreover, we have extended the partial-observation supervisor localization
to the case where inter-agent event communication is subject to bounded
and unbounded delay. To address communication delay, we have developed
an extended localization procedure based on explicit channel models and
relative coobservability. We have proved that the resulting local
controllers/preemptors collectively satisfy the communication delay
requirements. The above results are both illustrated by a timed workcell example.

In future research we shall extend the partial-observation localization procedure
to study distributed control of large-scale systems, by combing the proposed
supervisor localization with some efficient heterarchical synthesis procedure,
e.g. \cite{FenWon08}. We shall also study an alternative approach that
first synthesizes the full-observation centralized supervisor
and then performs localization to respect the observable event subsets specified {\it a priori}.


\small
\bibliographystyle{IEEEtran}
\bibliography{SCDES_Ref}


\end{document}